\documentclass{IEEEtran}
\usepackage{graphicx}
\usepackage{amsmath}
\usepackage{multicol}
\usepackage{multirow}
\usepackage{stfloats}
\usepackage{booktabs}
\usepackage{mathrsfs}
\usepackage{enumerate}
\usepackage{amssymb}
\usepackage{algorithm}
\usepackage{array}
\usepackage{underscore}
\usepackage{url}

\usepackage{hyperref}
\hypersetup{hidelinks}
\usepackage{flushend}

\usepackage{algorithmic}
\usepackage[square, comma, sort&compress, numbers]{natbib}
\usepackage{graphicx}
\usepackage{epstopdf}
\usepackage{caption}
\usepackage{diagbox}

\usepackage{caption}

\usepackage{subfigure}
\usepackage{color}

\usepackage[outerbars,color]{changebar}
\ifx\pdfoutput\undefined
\else\ifnum\pdfoutput>0
  \usepackage{pdfcolmk}
\fi\fi
\cbcolor{red}
\usepackage{fancyhdr}

\makeatletter
\newcommand{\Rmnum}[1]{\expandafter\@slowromancap\romannumeral #1@}
\makeatother

\usepackage{array}  \newcommand{\PreserveBackslash}[1]{\let\temp=\\#1\let\\=\temp}  \newcolumntype{C}[1]{>{\PreserveBackslash\centering}p{#1}}  \newcolumntype{R}[1]{>{\PreserveBackslash\raggedleft}p{#1}}  \newcolumntype{L}[1]{>{\PreserveBackslash\raggedright}p{#1}}

\ifCLASSINFOpdf
\else
\fi
\begin{document}
\title{\begin{Huge}Distributed Laser Charging: \\ A Wireless Power Transfer Approach \end{Huge}}

\author{Qingqing~Zhang,~\IEEEmembership{Student Member,~IEEE,}
Wen Fang,
Qingwen~Liu\IEEEauthorrefmark{1},~\IEEEmembership{Senior Member,~IEEE,}
Jun~Wu,~\IEEEmembership{Senior Member,~IEEE,}	
Pengfei~Xia,~\IEEEmembership{Senior Member,~IEEE,}	
and~Liuqing~Yang,~\IEEEmembership{Fellow,~IEEE}

	\thanks{Q.~Zhang, W. Fang, Q. Liu, J. Wu, and P. Xia, are with the College of Electronic and Information Engineering, Tongji University, Shanghai, China, (email: anne@tongji.edu.cn, wen.fang@tongji.edu.cn, qingwen.liu@gmail.com, wujun@tongji.edu.cn, pengfei.xia@gmail.com). }%
    \thanks{L. Yang is with the Department of Electrical and Computer Engineering, Colorado State University, Fort Collins, CO 80523, USA ({email:}{lqyang@engr.colostate.edu}).}%
\thanks{* Corresponding author.}
}

\maketitle

\begin{abstract}
Wireless power transfer (WPT) is a promising solution to provide convenient and perpetual energy supplies to electronics. Traditional WPT technologies face the challenge of providing Watt-level power over meter-level distance for Internet of Things (IoT) and mobile devices, such as sensors, controllers, smart-phones, laptops, etc.. Distributed laser charging (DLC), a new WPT alternative, has the potential to solve these problems and enable WPT with the similar experience as WiFi communications. In this paper, we present a multi-module DLC system model, in order to illustrate its physical fundamentals and mathematical formula. This analytical modeling enables the evaluation of power conversion or transmission for each individual module, considering the impacts of laser wavelength, transmission attenuation and photovoltaic-cell (PV-cell) temperature. Based on the linear approximation of electricity-to-laser and laser-to-electricity power conversion validated by measurement and simulation, we derive the maximum power transmission efficiency in closed-form. Thus, we demonstrate the variation of the maximum power transmission efficiency depending on the supply power at the transmitter, laser wavelength, transmission distance, and PV-cell temperature. Similar to the maximization of information transmission capacity in wireless information transfer (WIT), the maximization of the power transmission efficiency is equally important in WPT. Therefore, this work not only provides the insight of DLC in theory, but also offers the guideline of DLC system design in practice.

\end{abstract}


\IEEEpeerreviewmaketitle

\section{Introduction}\label{Section1}
Internet of Things (IoT) and mobile devices, such as sensors and smart-phones, are typically powered by batteries that have limited operation time. Sensors for IoT, especially sensors that being deployed in special environments such as volcanoes, are difficult to be charged. Meanwhile, carrying a power cord and looking for a power supplier to charge mobile devices incur great inconvenience. An alternative is thus to transfer power wirelessly, which virtually provides perpetual energy supplies. Hence, wireless power transfer (WPT) or wireless charging attracts great attention recently.

\begin{figure}
	\centering
	\includegraphics[scale=0.4]{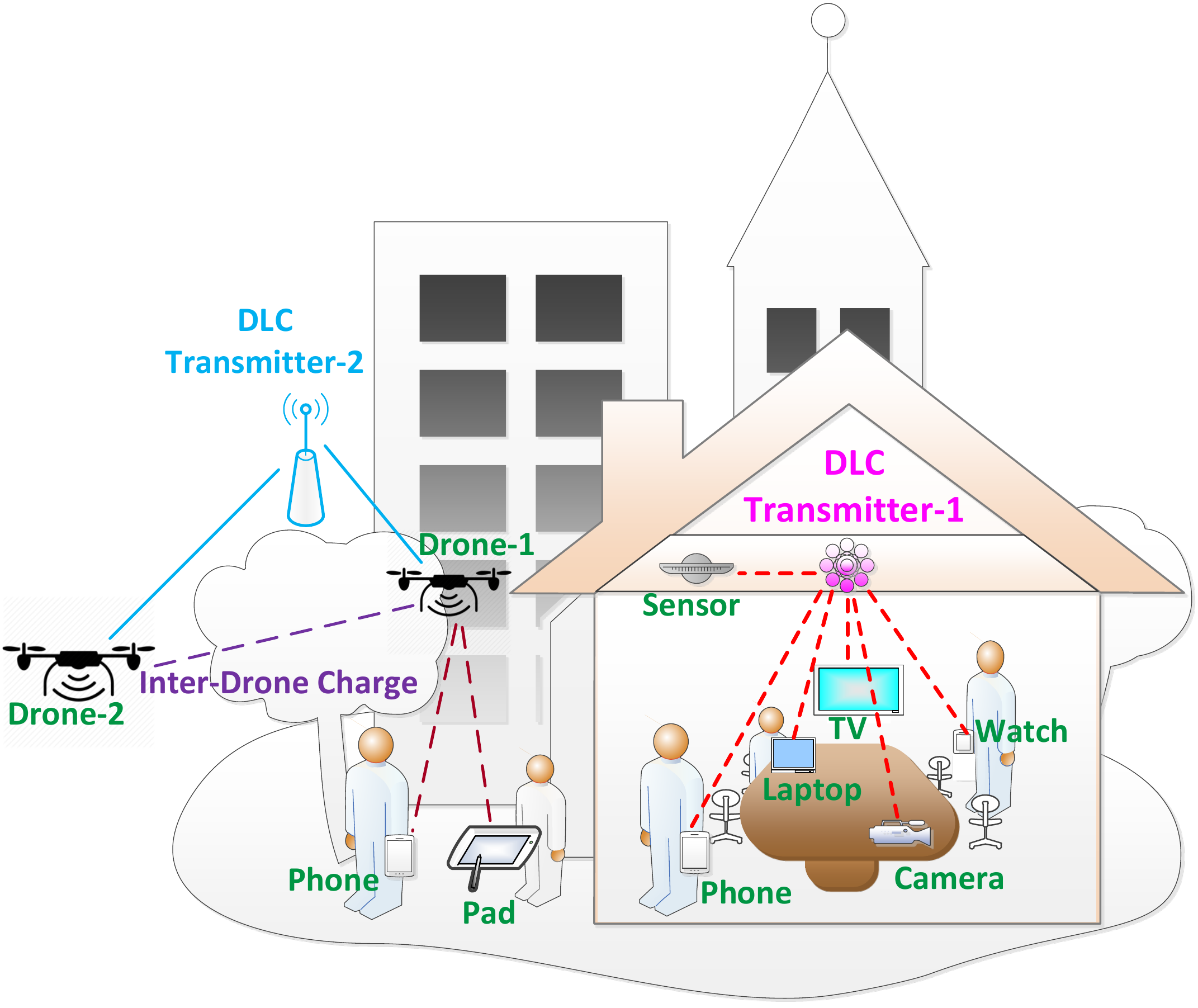}
	\caption{Distributed Laser Charging Applications}
	\label{adhocnetwork}
\end{figure}

Three major wireless charging technologies are surveyed in \cite{wirelesstechniques,electromagnetic}. Inductive coupling is safe and simple for implementation. However, it is limited by a short charging distance from a few millimeters to centimeters, which is only suitable for contact-charging devices like toothbrush. Magnetic resonance coupling has high charging efficiency. However, it is restricted by short charging distances and big coil sizes, which fits home appliances like TV. Electromagnetic (EM) radiation has long effective charging distances. However, it suffers from low charging efficiency and is unsafe when the EM power density exposure is high, hence is only favorable for low-power devices like sensors. In a nutshell, these traditional WPT technologies provide great wireless charging abilities for different application scenarios, whereas it is still challenging to offer sufficient power over long distance for safely charging IoT and mobile devices, e.g., smart-sensor, smart-phone, laptop, drone, etc., which usually need Watt-level power over meter-level distances.

To support the power and distance requirements for IoT and mobile devices, a distributed laser charging (DLC) system is presented in \cite{liu2016dlc}, which could transfer 2-Watt power over a 5-meter distance \cite{wi-charge}. By using inductive coupling or magnetic resonance coupling, IoT and mobile devices, say sensors and smart-phones, should typically be placed in a special charging cradle with a particular position. However, the DLC's self-aligning feature provides a more convenient way of charging IoT and mobile devices without specific positioning or tracking, as long as the transmitter and the receiver are in the line of sight (LOS) of each other. Different from EM radiation, DLC's wireless power transfer can be stopped immediately when this LOS is blocked by any object, which ensures the safety of DLC system. The size of the DLC receiver is sufficiently small to be embedded in a sensor or a smart-phone. The DLC transmitter can be installed on the ceiling like a lightbulb. In addition, multiple devices can be charged simultaneously by a single DLC transmitter \cite{niu2013optimal,anna2003opportunistic,zhou2004performance}. Therefore, DLC can provide IoT and mobile devices with safe WPT capability, which enables people to charge their devices with the similar experience as WiFi communications.

Fig.~\ref{adhocnetwork} illustrates the DLC potential applications. 
In Fig.~\ref{adhocnetwork}, in the room, DLC Transmitter-1 is combined with a light-emitting diode (LED) array and become a DLC-equipped lightbulb. Thus Transmitter-1 can be conveniently installed on the ceiling, and then provide wireless power to IoT and mobile devices within its coverage. In the outdoor scenario, Drone-1 is equipped with a DLC transmitter, which can charge IoT and mobile devices on demand. At the same time, a DLC receiver is also embedded in Drone-1. Thus, it can be remotely charged by DLC Transmitter-2, which acts as the power-supply base station on the ground. In addition, Drone-2 equipped with both DLC transmitter and receiver can play the role of a relay to receive power from DLC Transmitter-2 and transmit power to Drone-1 simultaneously.

Similar to the maximization of the \emph{information transmission capacity} of wireless channels in wireless information transfer (WIT), an important research topic in WPT is to maximize the \emph{power or energy transmission efficiency} \cite{zhangruimimo}. The wireless charging efficiency of a DLC system is affected by many factors, including laser wavelength, electricity-to-laser conversion efficiency, laser transmission attenuation, and laser-to-electricity conversion efficiency \cite{810nmtransmitter,1550nmtransmitter,laserenergy2009,green2015solar}. In this paper, we focus our study on the modeling of DLC system and its performance evaluation. In order to understand the fundamental mechanism of DLC system, we separate the DLC system into multiple conceptually independent modules. Thus, the corresponding power conversion or transmission for each module can be investigated individually, considering the impacts of laser wavelength, transmission attenuation, and photovoltaic-cell (PV-cell) temperature. Finally, the maximum power transmission efficiency in closed-form can be obtained from this modular analysis.

In this paper, a multi-module system model is proposed to describe the DLC system. The physical mechanism and mathematical formula are presented to describe the relationship between the stimulating electrical power and the output power, as well as the efficiency. The relationship between the supply power and the laser power, the relationship between the received laser power and the output power, and thus the relationship between the output power and the supply power are all depicted by both analytical results and illustrative graphs. The relationship between the electricity-to-laser conversion efficiency and the supply power, the relationship between the laser-to-electricity conversion efficiency and the received laser power, and thus the relationship between the maximum power transmission efficiency and the supply power are captured by closed-form expressions as well as being illustrated by figures. As a result, this work not only provides the insight of DLC in theory, but also offers the design guideline for DLC system implementation in practice.

In the rest of this paper, we will first review the DLC system and present the multi-module system model. Then, we will illustrate the analytical modeling of each module to investigate the corresponding working principles. After that, we will evaluate the performance of each module and derive the maximum DLC power transmission efficiency in closed-form. Finally, we will give summarizing remarks and discuss open issues for future research.


\section{DLC system}\label{Section2}
DLC is a WPT technology based on the distributed resonating laser presented in \cite{liu2016dlc}. Traditional laser systems belong to the scope of \emph{integrated resonating laser}, since all optical components are integrated in one single device. However, in DLC systems, the optical components are divided into two separate parts, the transmitter and the receiver, respectively. Therefore, the laser in DLC systems falls within the scope of \emph{distributed resonating laser}.

\begin{figure}
	\centering
	\includegraphics[scale=0.43]{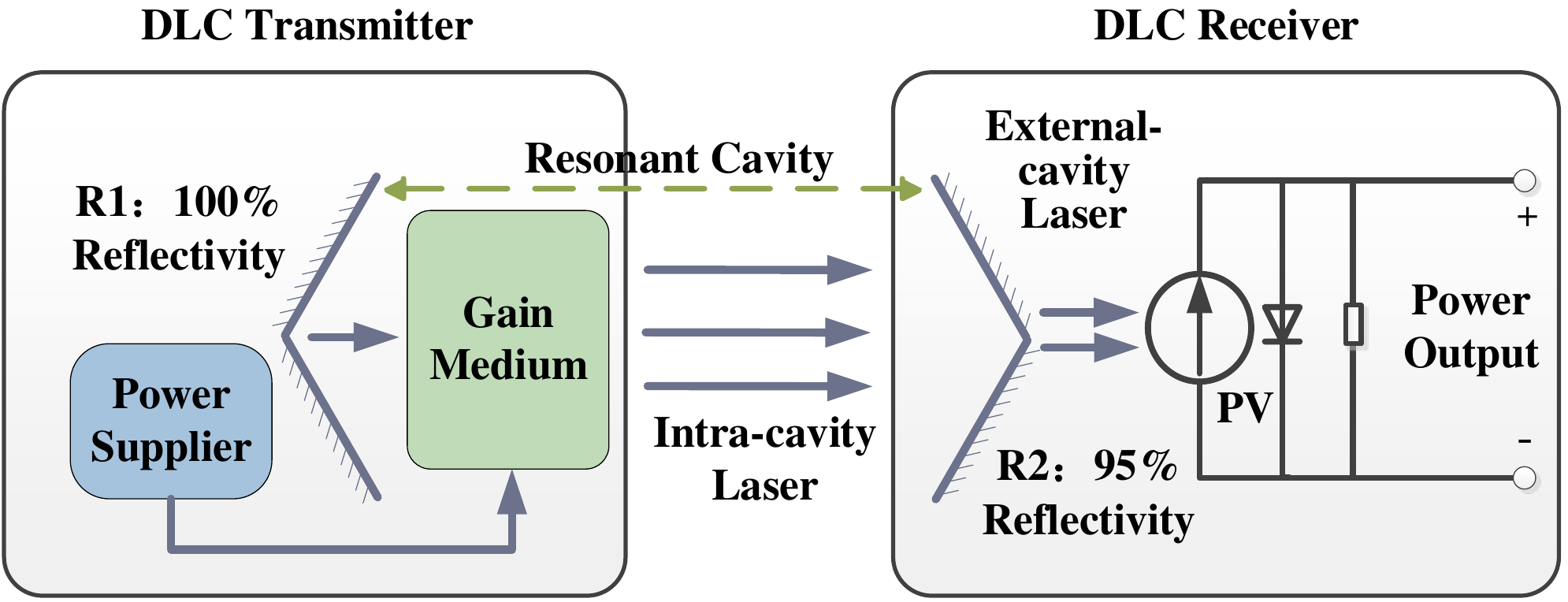}
	\caption{Distributed Laser Charging System Diagram}
	\label{adaptivepower}
\end{figure}

\begin{figure}
	\centering
    \includegraphics[scale=0.43]{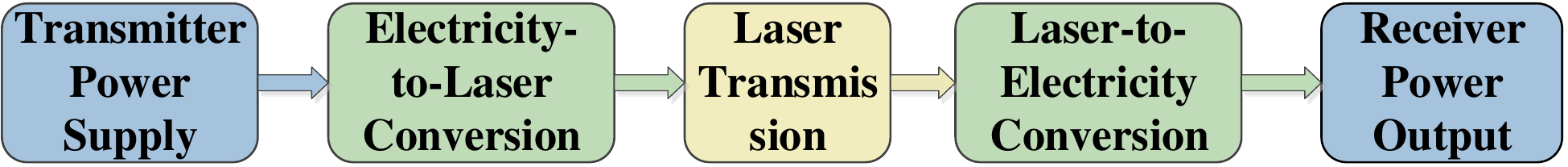}
	\caption{Distributed Laser Charging System Model}
    \label{effieiencyprocedure}
\end{figure}

Fig.~\ref{adaptivepower} shows the DLC system diagram described in \cite{liu2016dlc}. A retro-reflector mirror R1 with 100\% reflectivity and a gain medium are implemented at the transmitter. While in the receiver, a retro-reflector mirror R2 with exemplary 95\% partial reflectivity is contained. R1, R2 and the gain medium consist the resonant cavity, within which photons are amplified and form intra-cavity resonating laser. Photons that pass through R2 generates the external-cavity laser. The external-cavity laser power can be converted to electrical power by a photovoltaic-panel (PV-panel) installed behind mirror R2, which is similar to a solar panel. Fig.~\ref{adaptivepower} includes the power supplier at the transmitter and the power output at the receiver for the comprehensive DLC system design.

As specified in \cite{liu2016dlc}, in the DLC system, photons is amplified without concerning about the incident angle, as long as they travel along LOS of R1 and R2. Hence, the intra-cavity laser generated by the resonator can be self-aligned without specific positioning or tracking. This feature enables users to charge their devices without placing them in a specific position cautiously. Besides self-alignment, the DLC system is intrinsically-safe, since objects blocking the line-of-sight of intra-cavity laser can stop the laser immediately. These features offer DLC the capability of safely charging devices over long distance.

Fig. \ref{effieiencyprocedure} presents the system model to elaborate the wireless power transfer in the DLC system. This model illustrates a theoretical framework of power transfer by electricity-to-laser conversion, laser transmission, and laser-to-electricity conversion. The physical fundamentals and mathematical formulations of this modular model will be specified in the following section.


\section{Analytical Modeling}\label{Section3}
In this section, we will discuss each module of the DLC model in Fig.~\ref{effieiencyprocedure} and describe its wireless power transfer mechanism analytically. At the DLC transmitter, the power supplier provides electrical power to generate the intra-cavity laser. We will first introduce the electricity-to-laser conversion. Then, the intra-cavity laser will travel through the air and arrive at the DLC receiver. We will discuss the intra-cavity laser power attenuation along its transmission. At the DLC receiver, the intra-cavity laser will partially go through the mirror R2 and form the external-cavity laser, then the external-cavity laser will be converted into electricity by a PV-panel. We will analyze this laser-to-electricity conversion based on the PV engineering. Finally, the PV-panel output electrical power can be used to charge electronics. Based on the above analytical modeling, we will obtain the power conversion and transmission efficiency of each module and the overall power transmission efficiency.

\subsection{Electricity-to-Laser Conversion}\label{}
At the DLC transmitter, the electrical power $P_s$ is provided by the power supplier, which depends on the stimulating current $I_{t}$ and voltage $V_{t}$ as:
\begin{equation}\label{ps}
P_s = I_{t} V_{t}.
\end{equation}

The supply power $P_s$ can stimulate the gain medium to generate laser. Thus, the electrical power can be converted to the laser power.
We denote $P_l$ as the external-cavity laser power when the intra-cavity laser transmission efficiency is 100\%. It is well-known that laser can be generated, only when $I_{t}$ provided by the power supplier is over a certain threshold \cite{laserdiodes}. In the laser diode physics, the laser power $P_l$ relies on $I_{t}$ 
. Their relationship can be depicted as \cite{laserdiodes}:
\begin{equation}\label{powervscurrent}
P_{l}=\zeta \frac{h\upsilon}{q}(I_{t}-I_{th}),
\end{equation}
where $\zeta$ is the modified coefficient, $h$ is the Plunk constant, $\upsilon$ is the laser frequency, $q$ is the electronic charge constant, and $I_{th}$ is the current threshold.

Thus, the electricity-to-laser conversion efficiency $\eta_{el}$ can be figured out as:
\begin{equation}\label{etael}
  \eta_{el} = \frac{P_{l}}{P_s}.
\end{equation}

\subsection{Laser Transmission}\label{}
Laser power transmission attenuation means that laser power decreases along with its transmission through the air, which is similar to EM wave propagation power loss \cite{attenuation}. The laser power attenuation level depends on the transmission distance and air quality \cite{JMLiuphotonic,foghaze}. Relying on the above laser-generation mechanism, the intra-cavity laser can transmit from the transmitter to the receiver. During the transmission, laser may experience power attenuation. For simplicity, we assume that the laser diameter is a constant. This assumption could be validated by controlling aperture diameters of the DLC transmitter and receiver \cite{JMLiuphotonic}.

The laser transmission efficiency $\eta_{lt}$ can be modeled as \cite{JMLiuphotonic}:
\begin{equation}\label{etalt}
  \eta_{lt}=\frac{P_{r}}{P_l}= e^{-\alpha d},
\end{equation}
where $P_{r}$ is the external-cavity laser power received at the DLC receiver, $\alpha$ is the laser attenuation coefficient, and $d$ is the distance. When $d$ is close to zero, the laser transmission efficiency approaches 100\%. In this situation, $P_{r}$ is approximate to $P_{l}$.

$\alpha$ can be depicted as:
\begin{equation}\label{alphac}
  \alpha = \frac{\sigma}{\kappa} \Big(\frac{\lambda}{\chi}\Big)^{-\rho},
\end{equation}
where $\sigma$ and $\chi$ are two constants, $\kappa$ is the visibility, $\lambda$ is the wavelength, and $\rho$ is the size distribution of the scattering particles. $\rho$ depends on visibility, which will be discussed later.

\subsection{Laser-to-Electricity Conversion}\label{}
\begin{figure}
	\centering
	\includegraphics[scale=0.8]{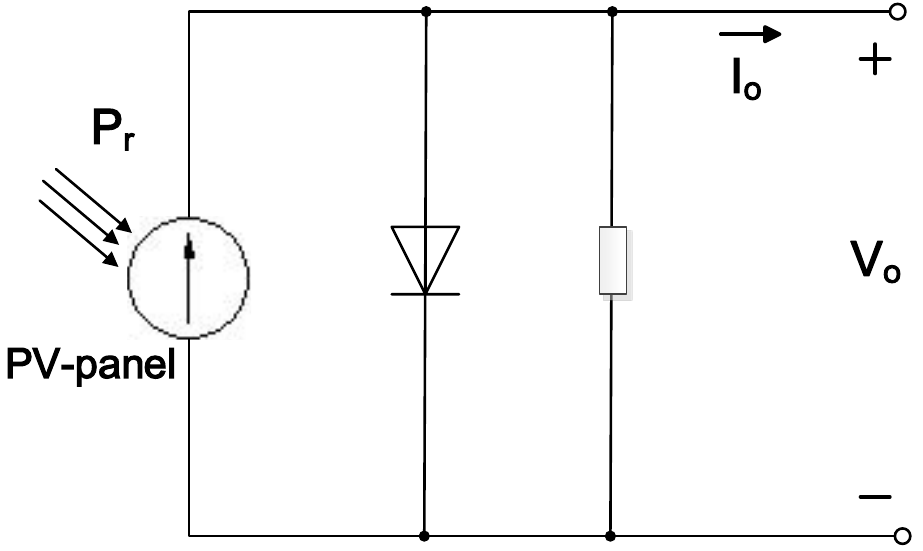}
	\caption{PV-panel Power Conversion Circuit Model}
	\label{singlediodeonly}
\end{figure}

\begin{table}[b]
\centering
\caption{Transmission or Conversion Efficiency}
\begin{tabular}{C{6.0cm} C{1.5cm}}
\hline
 \textbf{Parameter} & \textbf{Symbol}  \\
\hline
\bfseries{Electricity-to-laser conversion efficiency} & {${\eta}_{el}$} \\
\bfseries{Laser transmission efficiency} & {${\eta}_{lt}$} \\
\bfseries{Laser-to-electricity conversion efficiency} & {${\eta}_{le}$} \\
\hline
\bfseries{The overall DLC power transmission efficiency} & {${\eta}_{o}$} \\
\hline
\label{convertefficiency}
\end{tabular}
\end{table}

\begin{figure}
	\centering
    \includegraphics[scale=0.6]{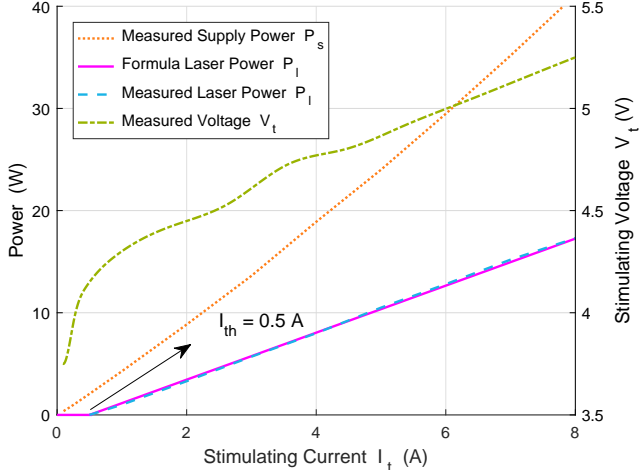}
	\caption{Electricity-to-Laser Conversion Power, Voltage and Current (810nm)}
    \label{810ptI}
\end{figure}

At the DLC receiver, the external-cavity laser power can be converted to electrical power. To illustrate the laser-to-electricity conversion mechanism, the single-diode equivalent circuit model of a PV-panel is depicted in Fig.~\ref{singlediodeonly} \cite{secondauthorPV}. The PV-panel output voltage $V_{o}$, and current $I_{o}$ can be characterized as \cite{secondauthorPV}:
\begin{equation}\label{Io}
  I_{o}=I_{sc} - I_{s}(e^{V_{o}/V_m}-1),
\end{equation}
where $I_{sc}$ is the PV-panel short-circuit current, $I_s$ is the saturation current, i.e., the diode leakage current density in the absence of light, and $V_m$ is the ``thermal voltage'', which can be defined as:
\begin{equation}\label{Vm}
  V_m=\frac{nkT}{q},
\end{equation}
where $n$ is the PV-panel ideality factor, $k$ is the Boltzmann constant, and $T$ is the absolute PV-cell temperature. Then, the PV-panel output power $P_{o}$, which relies on $I_{o}$ and $V_{o}$, can be obtained as:
\begin{equation}\label{po}
  P_{o}=I_{o}V_{o}.
\end{equation}

Therefore, the laser-to-electricity conversion efficiency, i.e. the PV-panel conversion efficiency, $\eta_{le}$ depends on $P_{o}$ and $P_{r}$, which can be depicted as:
\begin{equation}\label{etapv}
{\eta}_{le}=\frac{P_{o}}{P_{r}}=\frac{I_{o}V_{o}}{P_{r}}.
\end{equation}

In summary, the PV-panel converts the received laser power $P_{r}$ to the output power $P_{o}$ with the efficiency $\eta_{le}$.

\subsection{DLC Power Transmission Efficiency}\label{}
Based on the above analysis for each individual module of the DLC system model, the DLC power transmission efficiency from the power supplier at the transmitter to the power output at the receiver can be depicted as:
\begin{equation}\label{etao}
  {\eta}_{o}={\eta}_{el} {\eta}_{lt} {\eta}_{le}.
\end{equation}
The conversion or transmission efficiency of each module and the DLC power transmission efficiency are listed in Table \ref{convertefficiency}.

The numerical evaluation of the DLC system model will be presented in the next section.


\section{Numerical Evaluation}\label{Section4}
Based on the analytical modeling in the previous section, we can find that the DLC system efficiency varies with laser wavelength, transmission attenuation and PV-cell temperature. Their impacts on the performance of each module as well as the overall DLC system will be discussed in this section. The numerical evaluation is implemented in MATLAB and Simulink.

\subsection{Electricity-to-Laser Conversion}\label{}

\begin{table}[b]
\newcommand{\tabincell}[2]{\begin{tabular}{@{}#1@{}}#2\end{tabular}}
\centering
\caption{Electricity-to-Laser Conversion Parameters}
\begin{tabular}{C{3.4cm} C{0.1cm} C{4.0cm}}
\hline
 \textbf{Parameter} & \textbf{Symbol} & \tabincell{c}{\textbf{Value} \\ \textbf{810nm} \qquad \textbf{1550nm}} \\
\hline
\bfseries{Boltzmann constant}                                    & {k}           & $1.38064852\times10^{-23} J/K$ \\
\bfseries{Planck constant}                                       & {h}           & $6.62606957\times10^{-34} J\cdot s$ \\
\bfseries{Electronic charge constant}                                       & {q}           & $1.6\times10^{-19} C$ \\
\bfseries{Laser wavelength }                                     & {$\lambda$}   & \tabincell{c}{$810nm$ \qquad $1550nm$}  \\
\bfseries{Laser frequency}       & {$\upsilon$}  & \tabincell{c}{$3.7\times10^{14} Hz$ \qquad $1.9\times10^{14} Hz$} \\
\bfseries{Stimulation current threshold}                         & {I$_{th}$}    & \tabincell{c}{$0.5 A$ \qquad $0.6 A$}\\
\bfseries{Modified coefficient}                                  & {$\zeta$}     & \tabincell{c}{$1.5$ \qquad $3.52$} \\
\bfseries{$\textbf{P}_{\textbf{l}}$-$\textbf{P}_{\textbf{s}}$ curve fitting parameter} & {$a_1$} & \tabincell{c}{$0.445$ \qquad $0.34$} \\
\bfseries{$\textbf{P}_{\textbf{l}}$-$\textbf{P}_{\textbf{s}}$ curve fitting parameter} & {$b_1$} & \tabincell{c}{$-0.75$ \qquad $-1.1$} \\
\hline
\label{paramaters}
\end{tabular}
\end{table}

\begin{figure}
	\centering
    \includegraphics[scale=0.6]{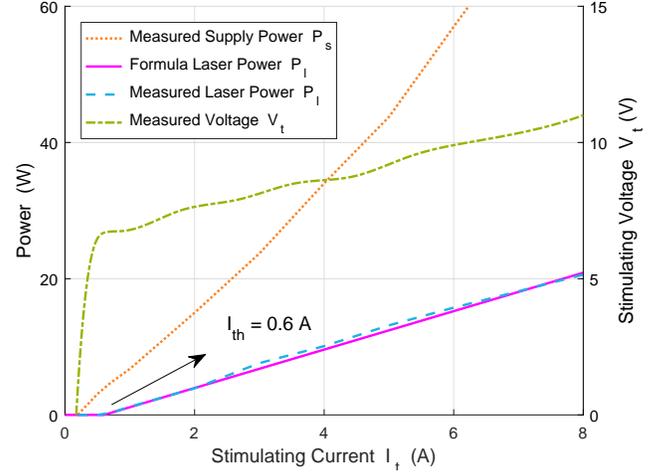}
	\caption{Electricity-to-Laser Conversion Power, Voltage and Current (1550nm)}
    \label{1550ptI}
\end{figure}

Electrical supply power $P_{s}$ provided by the power supplier at the transmitter depending on the stimulating current $I_{t}$ and voltage $V_{t}$, as in \eqref{ps}. Based on the measurement of $I_{t}$, $V_{t}$, and thus $P_{s}$, for the laser systems ($\lambda$ is 800-820nm and 1540-1560nm, respectively) in \cite{810nmtransmitter,1550nmtransmitter}, the measured supply power $P_{s}$, the measured laser power $P_{l}$, the stimulating current $I_{t}$, and the stimulating voltage $V_{t}$ are shown for 810nm and 1550nm in Fig.~\ref{810ptI} and Fig.~\ref{1550ptI}, respectively. From the dashed-lines for the measured laser power in Fig.~\ref{810ptI} and Fig.~\ref{1550ptI}, the modified coefficient $\zeta$ in \eqref{powervscurrent} can be determined and listed in Table \ref{paramaters}. Thus, from \eqref{powervscurrent}, the formulated laser power curves are given as the solid-lines in Fig.~\ref{810ptI} and Fig.~\ref{1550ptI}, respectively.

\begin{figure}
	\centering
    \includegraphics[scale=0.6]{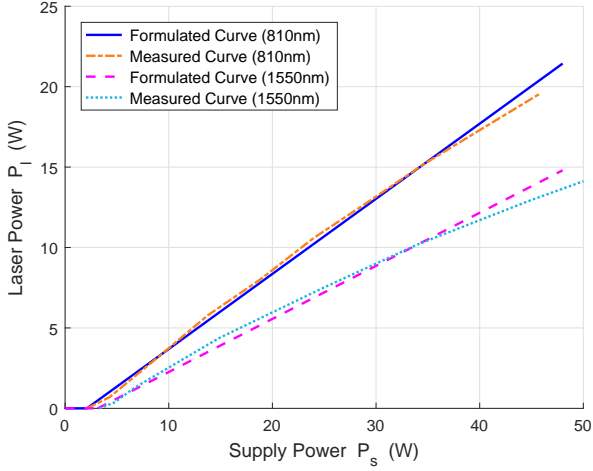}
	\caption{Laser Power vs. Supply Power \protect\\ \qquad}
    \label{8101550plps}
\end{figure}

In Fig.~\ref{810ptI} and Fig.~\ref{1550ptI}, the relationship between $P_{l}$ and $P_{s}$ is illustrated in Fig.~\ref{8101550plps}. We adopt the linear formula to approximate this power conversion as:
\begin{equation}\label{plps}
P_{l} \approx a_{1} P_{s}+b_{1}.
\end{equation}
The measured and formulated curves in Fig.~\ref{8101550plps} depict the linear approximation between $P_{l}$ and $P_{s}$ based on \eqref{plps}, when the wavelength $\lambda$ is about 810nm and 1550nm, respectively. We can find that the fitting curves match the measurement very well in the given supply power and laser power range in Fig.~\ref{8101550plps}.

From \eqref{etael} and \eqref{plps}, we can obtain the electricity-to-laser conversion efficiency $\eta_{el}$ as:
\begin{equation}\label{etaelall}
\eta_{el}= \frac{P_{l}}{P_{s}} = a_{1}+\frac{b_{1}}{P_{s}}.
\end{equation}

\begin{figure}
	\centering
    \includegraphics[scale=0.6]{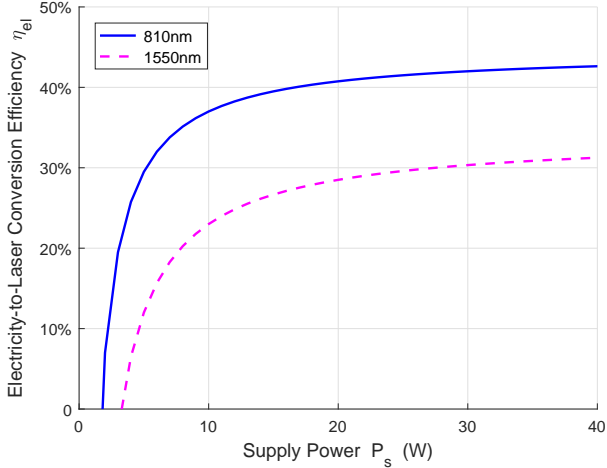}
	\caption{Electricity-to-Laser Conversion Efficiency vs. Supply Power}
    \label{8101550etael}
\end{figure}

The solid-line and dashed-line in Fig.~\ref{8101550etael} illustrate $\eta_{el}$ for 810nm and 1550nm, respectively. The initial $P_{s}$ supply power threshold in Fig.~\ref{8101550etael} is corresponding to the current threshold $I_{th}$ for $P_{l}$ in Fig.~\ref{810ptI} and Fig.~\ref{1550ptI}. In Fig.~\ref{8101550etael}, $\eta_{el}$ starts to increase dramatically from the initial supply power $P_{s}$ threshold and will reach the plateau as $P_{s}$ increases. The plateau of $\eta_{el}$ for 810nm laser is around 43\%, which is higher than 31\% for 1550nm laser.

\subsection{Laser Transmission}\label{}
From \eqref{etalt} and \eqref{alphac}, the laser power attenuation coefficient in transmission can be determined under three typical scenarios, i.e., clear air, haze, and fog. For the three scenarios, the size distribution of the scattering particles $\rho$ in \eqref{alphac} can be specified as \cite{foghaze}:
\begin{equation}\label{rhoc}
\rho = \left\{
             \begin{array}{lr}
             1.3 \qquad\qquad\qquad\  \mathrm{for \  clear \  air} \quad\  (6km\leq \kappa \leq 50km), &  \\
             0.16 \kappa+0.34 \qquad\ \mathrm{for \  haze} \qquad\quad (1km\leq \kappa \leq 6km),\\
             0 \qquad\qquad\qquad\quad\  \mathrm{for \  fog} \qquad\quad\ \  (\kappa \leq 0.5km), &
             \end{array}
\right.
\end{equation}
where $\kappa$ is the visibility.

\begin{figure}
	\centering
    \includegraphics[scale=0.6]{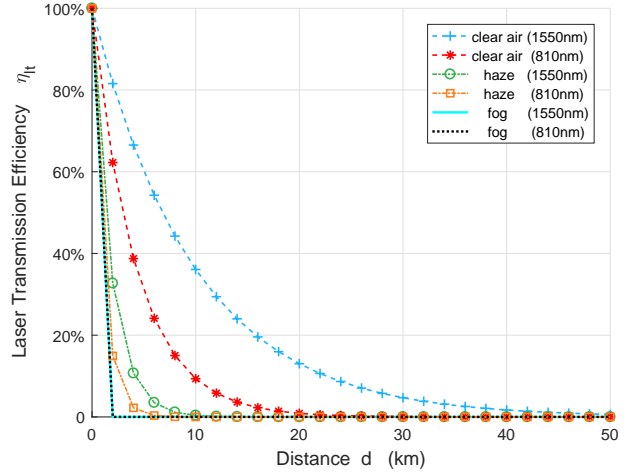}
	\caption{Laser Transmission Efficiency vs. Distance}
    \label{transmittance}
\end{figure}

\begin{table}[b]
\newcommand{\tabincell}[2]{\begin{tabular}{@{}#1@{}}#2\end{tabular}}
\centering
\caption{Laser Transmission Parameters}
\begin{tabular}{C{1.0cm} C{7.0cm}}
\hline
 \textbf{Parameter} &\tabincell{c}{\textbf{Value} \\ \quad\quad \textbf{Clear Air} \qquad\qquad\quad \textbf{Haze} \qquad\qquad\qquad \textbf{Fog}} \\
\hline
\bfseries{$\sigma$} & {\quad\qquad $3.92$} \\
\bfseries{$\chi$}   & \quad\qquad $550nm$ \\
\bfseries{$\kappa$} & {\quad\qquad \  $10km$ \qquad\qquad\quad $3km$ \qquad\qquad\quad\  $0.4km$} \\
\bfseries{$\rho$}   & {\quad\qquad $1.3$ \qquad\qquad $0.16\kappa+0.34$ \qquad\qquad $0$} \\
\hline
\label{pathloss}
\end{tabular}
\end{table}

Along with $\rho$, the other attenuation parameters are listed in Table \ref{pathloss}. Thus, the relationship between $\eta_{lt}$ and the transmission distance $d$ can be obtained from \eqref{etalt} and \eqref{alphac}, which is illustrated in Fig.~\ref{transmittance}. It is clear that $\eta_{lt}$ decays exponentially to zero as $d$ increases. Meanwhile, for the same laser wavelength, laser power attenuation depends on the visibility $\kappa$. Laser power attenuation increases when $\kappa$ decreases. As can be seen in Fig.~\ref{transmittance}, for clear air, haze and fog, given the same $d$, the laser power attenuation for short-wavelength is more than that of long-wavelength. For clear air and haze, laser attenuation for 810nm is much more than that of 1550nm. However, for fog, since $\rho$ takes 0 for both 810nm and 1550nm, the coefficient $\alpha$ has the same value. Therefore, the laser attenuation in fog does not dependent on $\lambda$.

\begin{figure}
	\centering
    \includegraphics[scale=0.6]{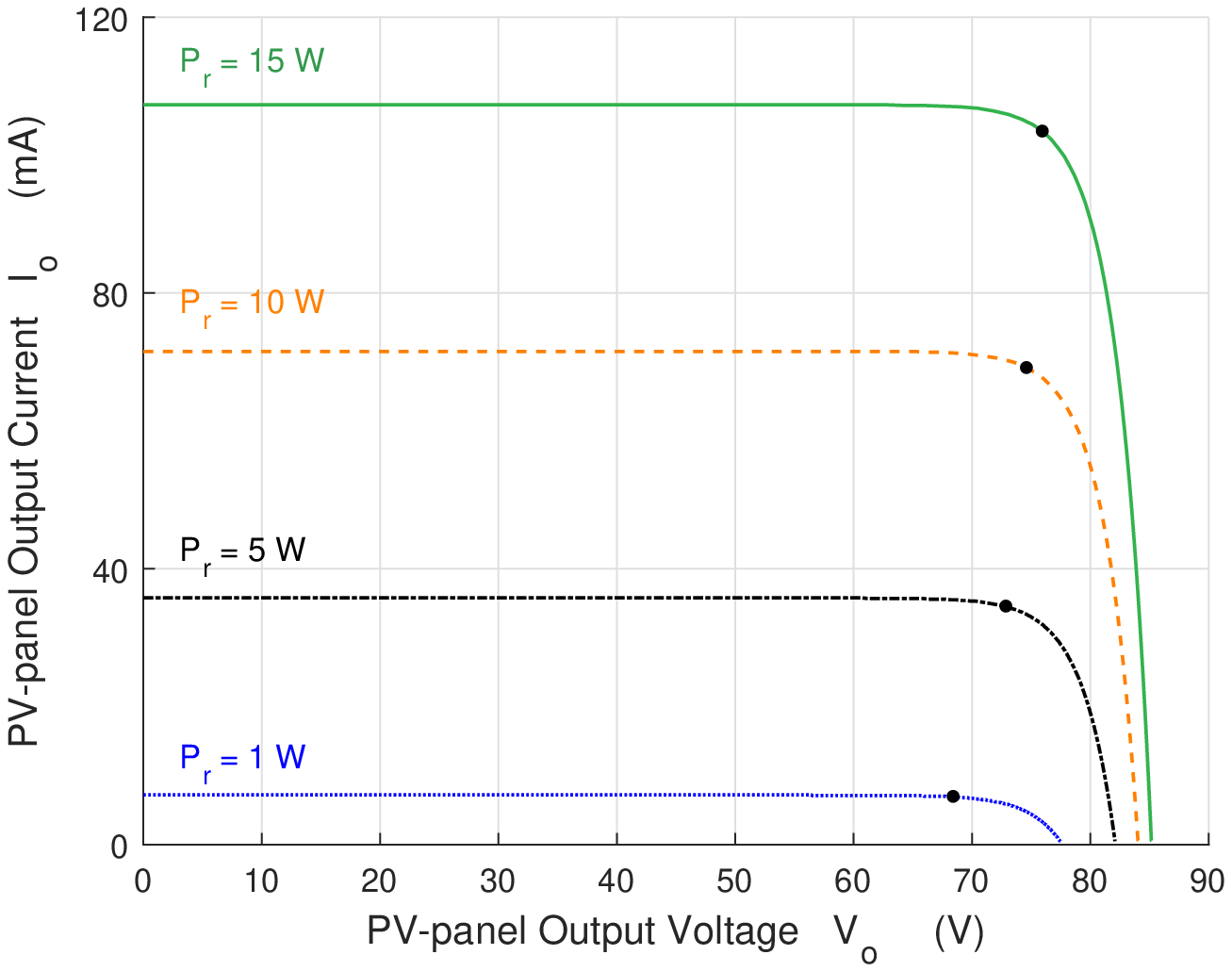}
	\caption{PV-panel Output Current vs. Voltage ($\lambda$ = 810nm)}
    \label{810Iirradiance}
\end{figure}

\begin{figure}
	\centering
    \includegraphics[scale=0.6]{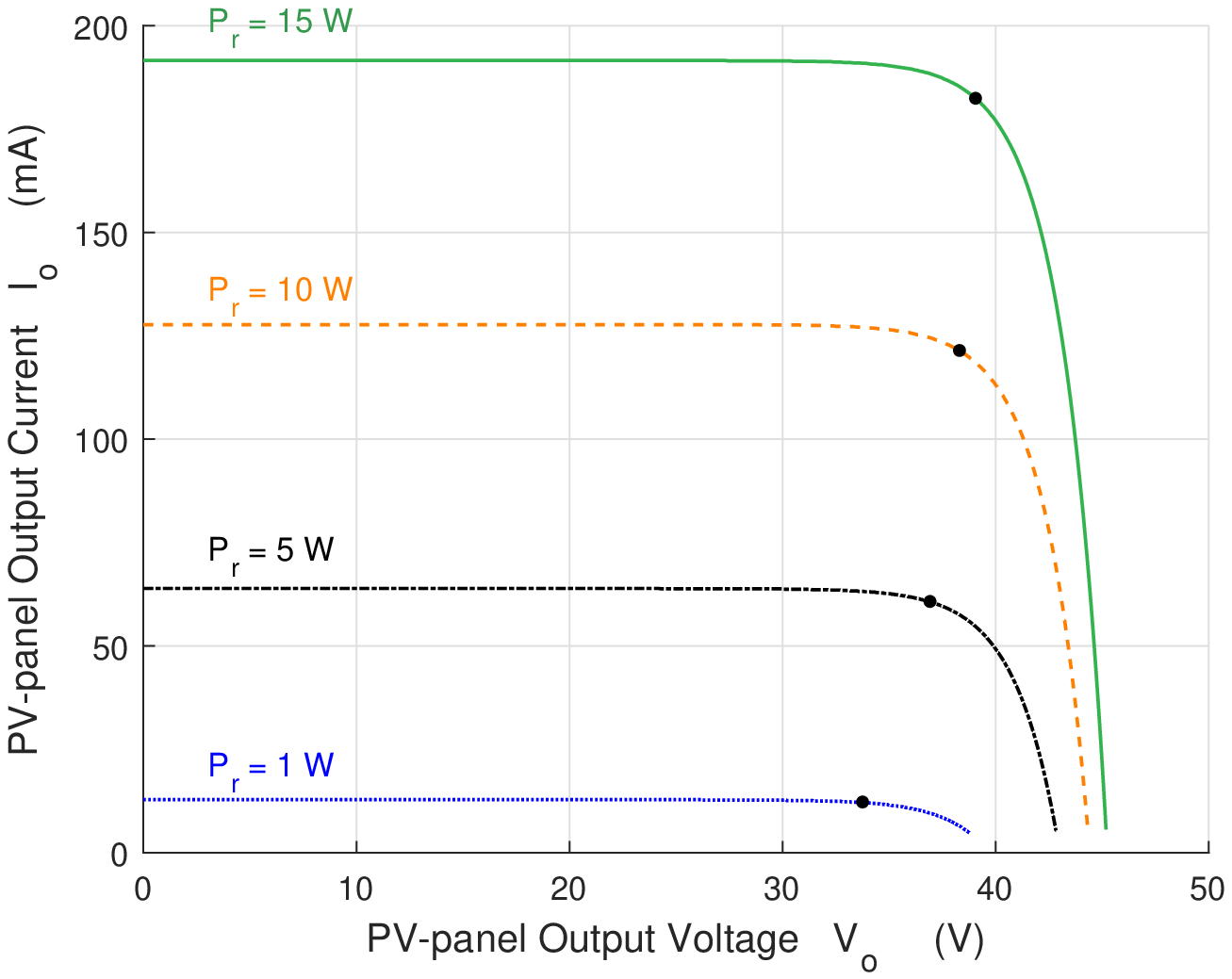}
	\caption{PV-panel Output Current vs. Voltage ($\lambda$ = 1550nm)}
    \label{1550Iirradiance}
\end{figure}

\begin{figure}
	\centering
    \includegraphics[scale=0.6]{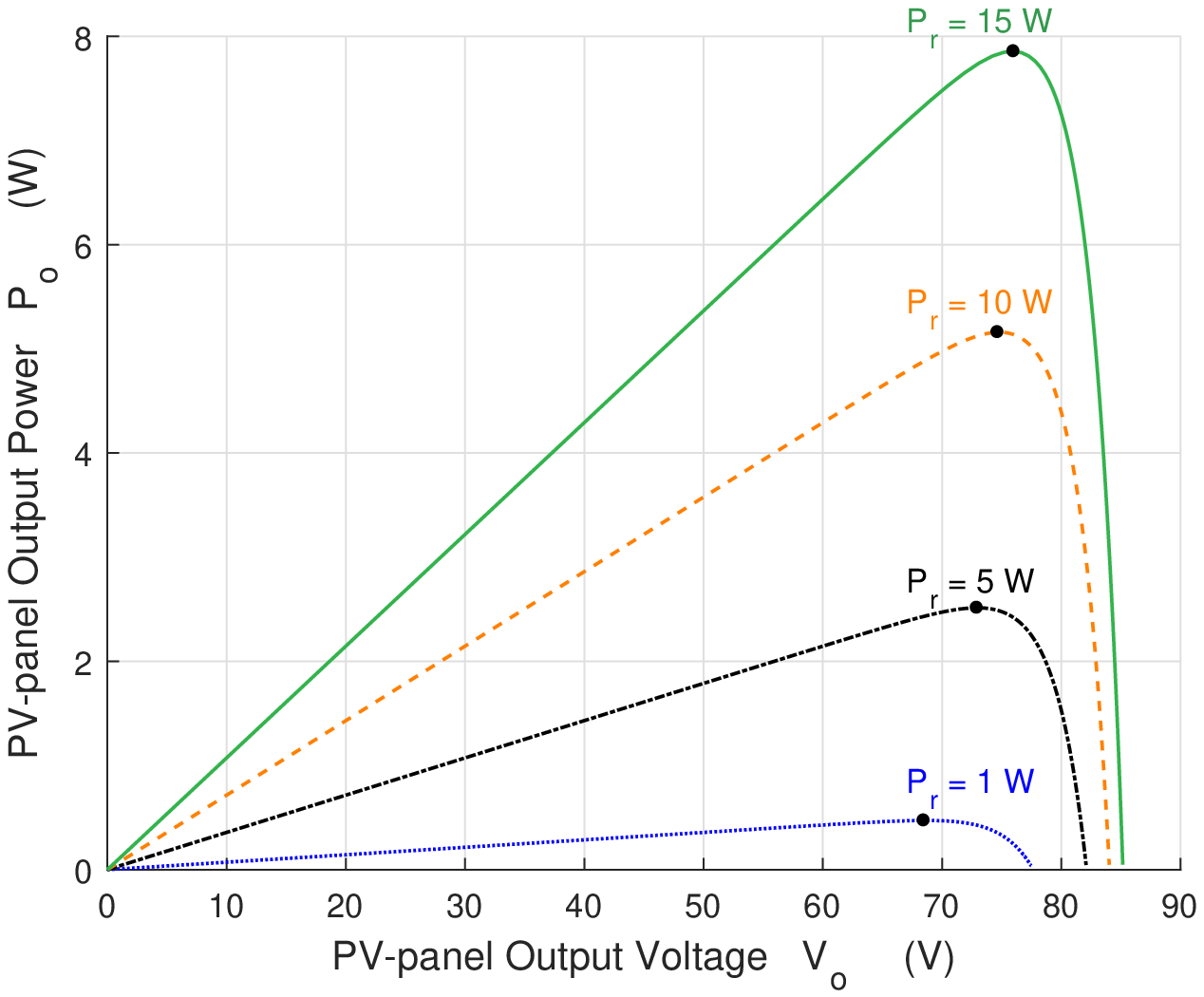}
	\caption{PV-panel Output Power vs. Voltage ($\lambda$ = 810nm)}
    \label{810Pirradiance}
\end{figure}

\begin{table}
\newcommand{\tabincell}[2]{\begin{tabular}{@{}#1@{}}#2\end{tabular}}
\centering
\caption{Laser-to-Electricity Conversion Parameters}
\begin{tabular}{C{1.7cm} C{0.3cm} C{4.5cm}}
\hline
 \textbf{Parameter} & \textbf{Symbol} & \tabincell{c}{\textbf{Value} \\ \textbf{810nm} \qquad \textbf{1550nm}} \\
\hline
\bfseries{Short-circuit current}                                & {$I_{sc}$}    & \tabincell{c}{$0.16732 A$ \qquad $0.305 A$\qquad\ }\\
\bfseries{Open-circuit voltage}                                 & {$V_{oc}$}    & \tabincell{c}{$1.2 V$ \qquad $0.464 V$} \\
\bfseries{Irradiance used for measurement}                 & {$I_{r0}$}    & \tabincell{c}{$36.5 W/cm^2$ \qquad\qquad $2.7187 W/cm^2$} \\
\bfseries{Laser frequency}       & {$\upsilon$}  & \tabincell{c}{$3.7037\times10^{14} Hz$ \quad $1.9355\times10^{14} Hz$} \\
\bfseries{Quality factor}                                       & {$n$}            & \tabincell{c}{$1.5$ \qquad\quad $1.1\qquad$} \\
\bfseries{Number of series cells}                               & {$N$}           & $72$ \\
\bfseries{PV-panel material }                                   & { }   & \tabincell{c}{GaAs-based \qquad GaSb-based}  \\
\bfseries{Measurement temperature}                              & {$T$}       & \tabincell{c}{$25^{\circ}C$ \qquad\  $120^{\circ}C$} \\
\bfseries{Simulation temperature}    & {}     & \tabincell{c}{$0^{\circ}C$ / $25^{\circ}C$ / $50^{\circ}C$} \\
\bfseries{$\textbf{P}_{\textbf{m}}$-$\textbf{P}_{\textbf{r}}$ curve fitting parameter}     & {$a_2$}       & \tabincell{c}{0.546/0.541/0.537 \ 0.543/0.498/0.453} \\
\bfseries{$\textbf{P}_{\textbf{m}}$-$\textbf{P}_{\textbf{r}}$ curve fitting parameter}   & {$b_2$}   & \tabincell{c}{-0.213/-0.231/-0.249 \ -0.276/-0.299/-0.321} \\
\hline
\label{pvparamaters}
\end{tabular}
\end{table}

\subsection{Laser-to-Electricity Conversion}\label{}

\begin{figure}
	\centering
    \includegraphics[scale=0.6]{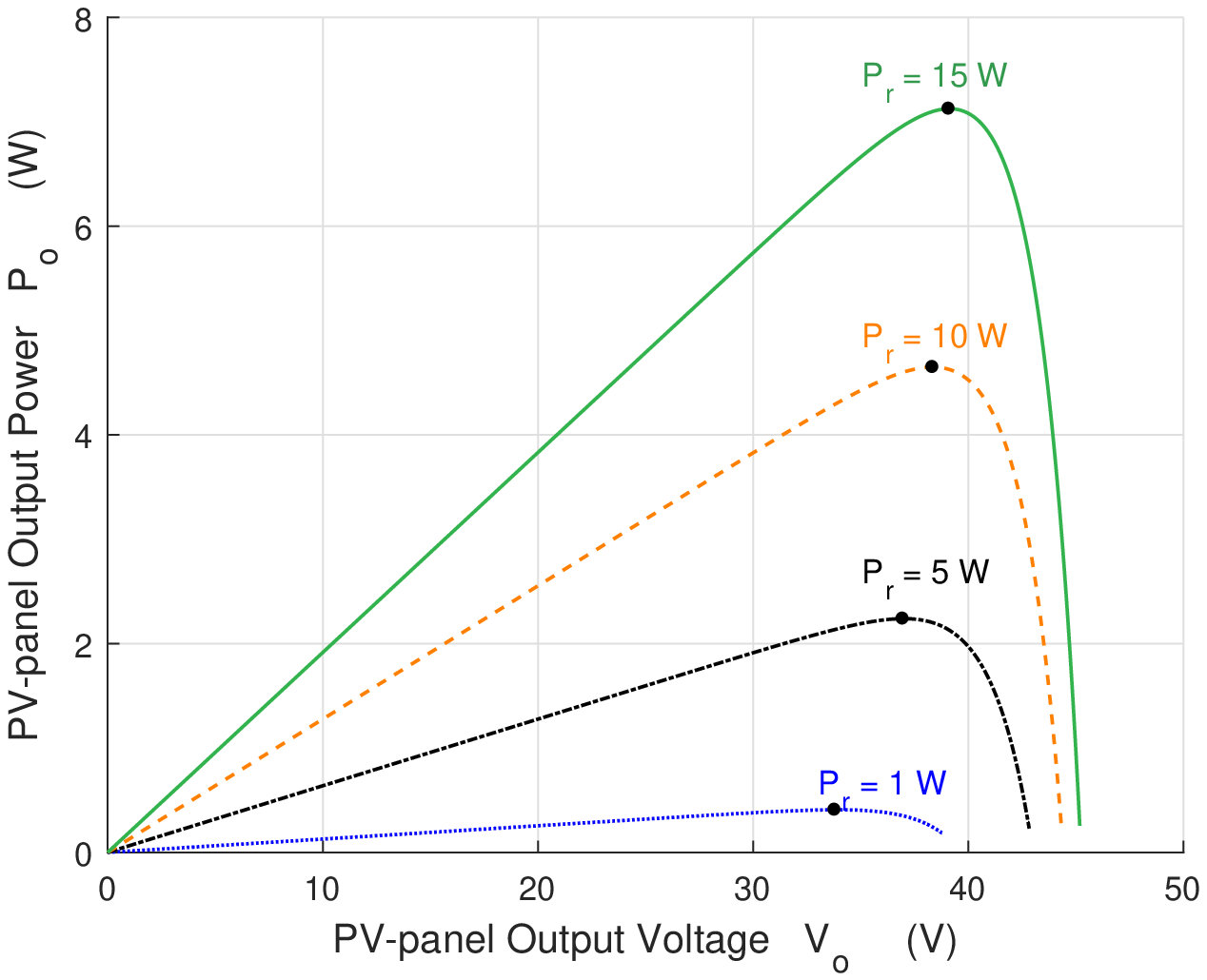}
	\caption{PV-panel Output Power vs. Voltage ($\lambda$ = 1550nm)}
    \label{1550Pirradiance}
	\centering
    \includegraphics[scale=0.6]{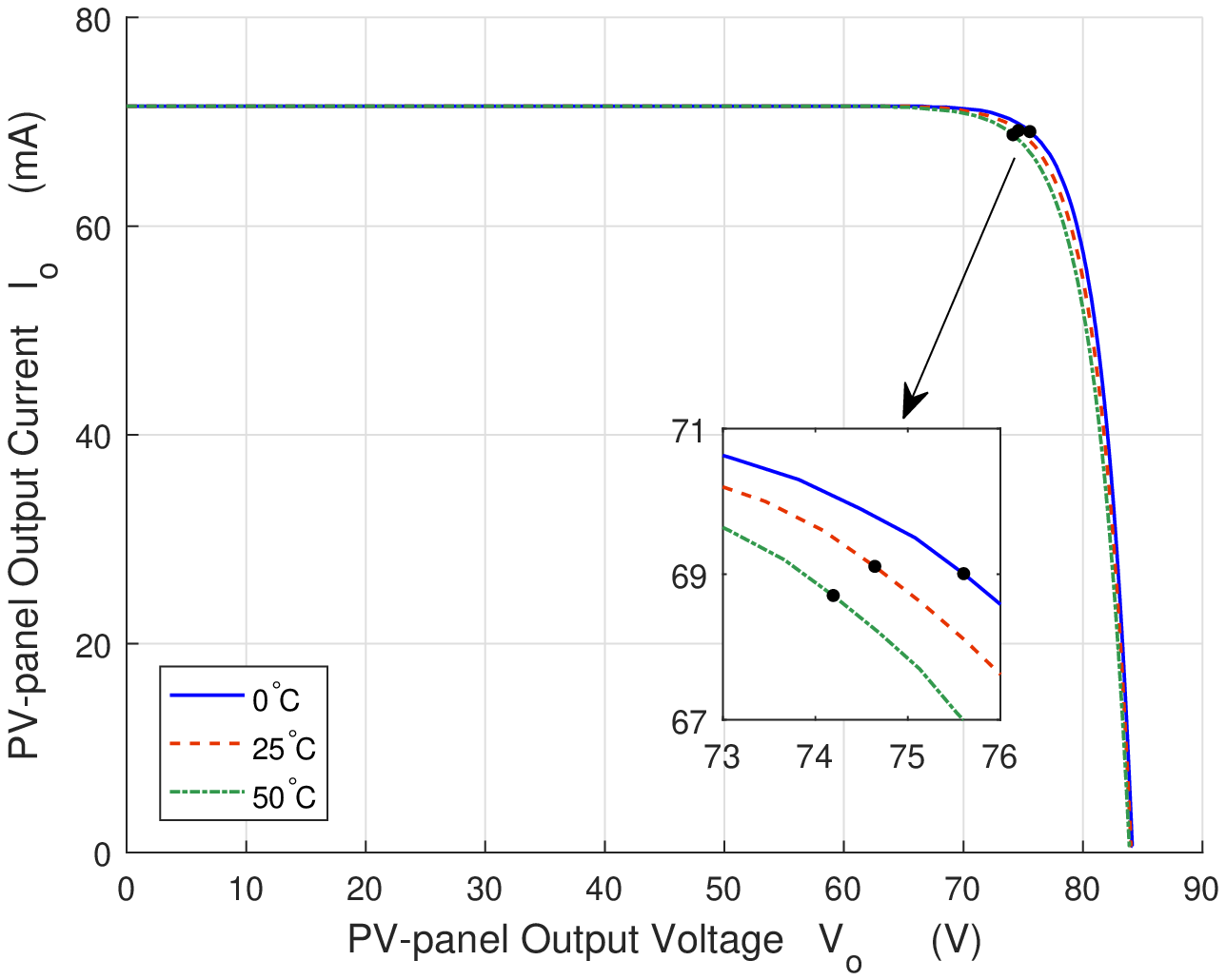}
	\caption{PV-panel Output Current vs. Voltage ($\lambda$ = 810nm)}
    \label{810Itemperature}
	\centering
    \includegraphics[scale=0.6]{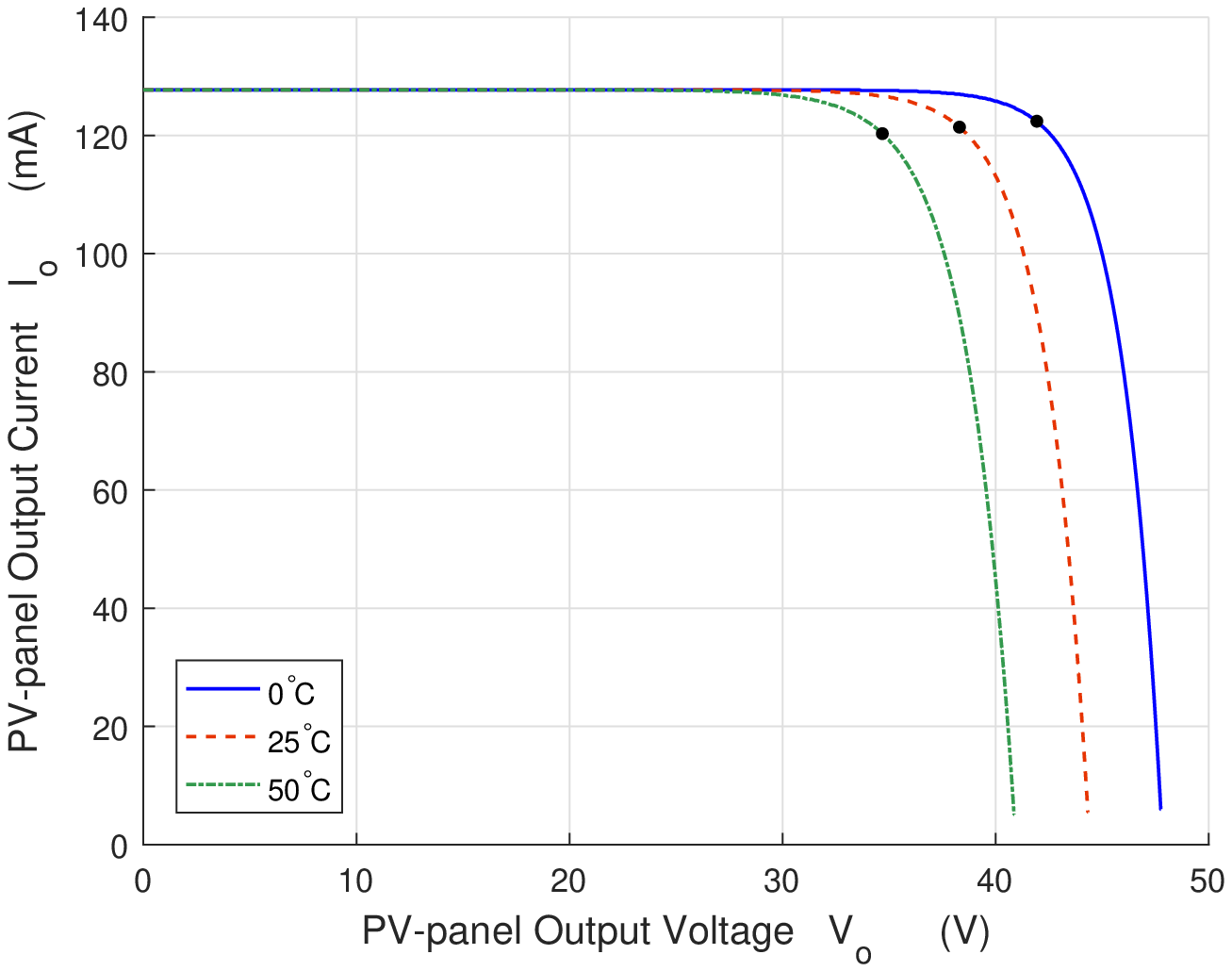}
	\caption{PV-panel Output Current vs. Voltage ($\lambda$ = 1550nm)}
    \label{1550Itemperature}
\end{figure}

\begin{figure}
	\centering
    \includegraphics[scale=0.6]{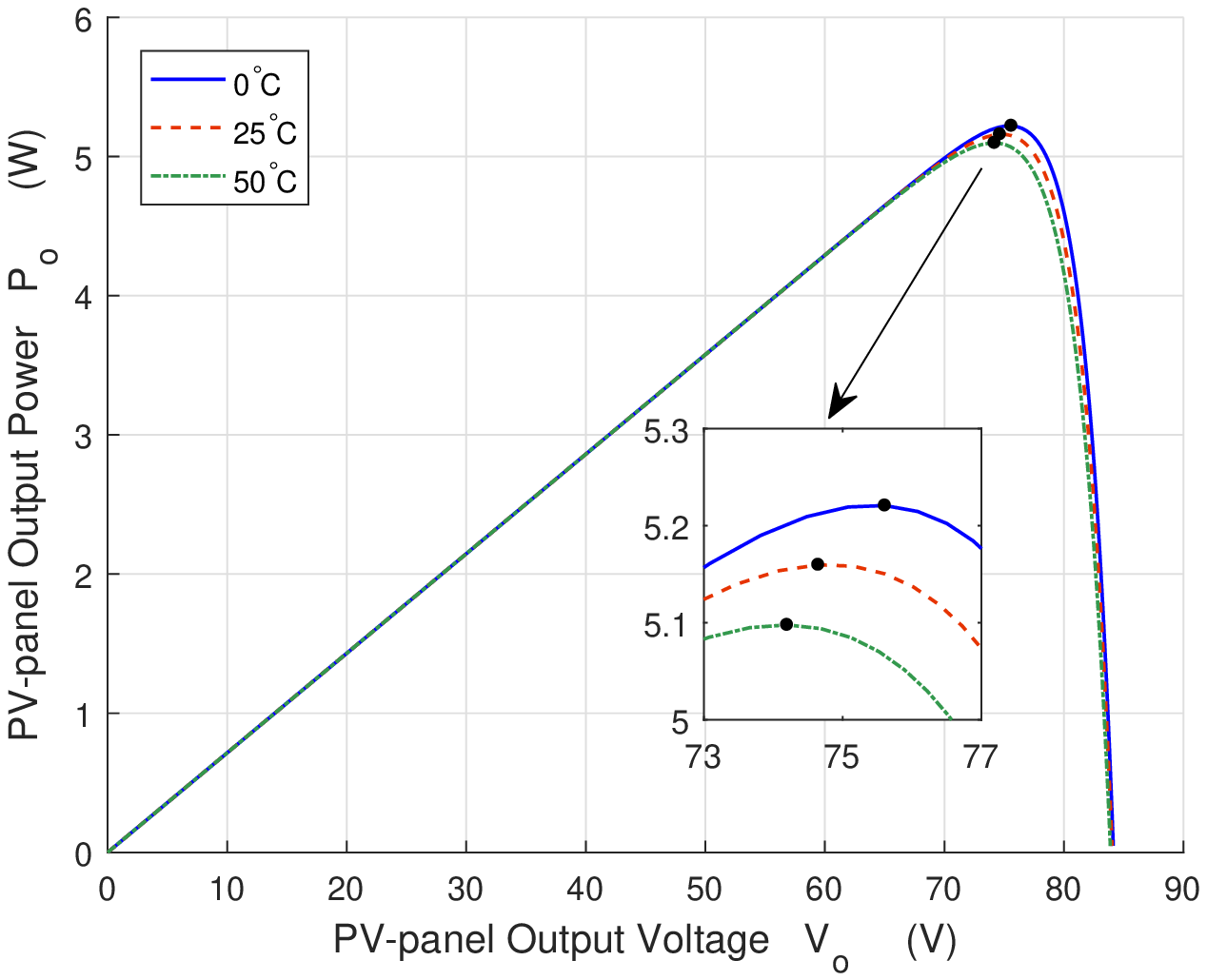}
	\caption{PV-panel Output Power vs. Voltage ($\lambda$ = 810nm)}
    \label{810Ptemperature}
%
	\centering
    \includegraphics[scale=0.6]{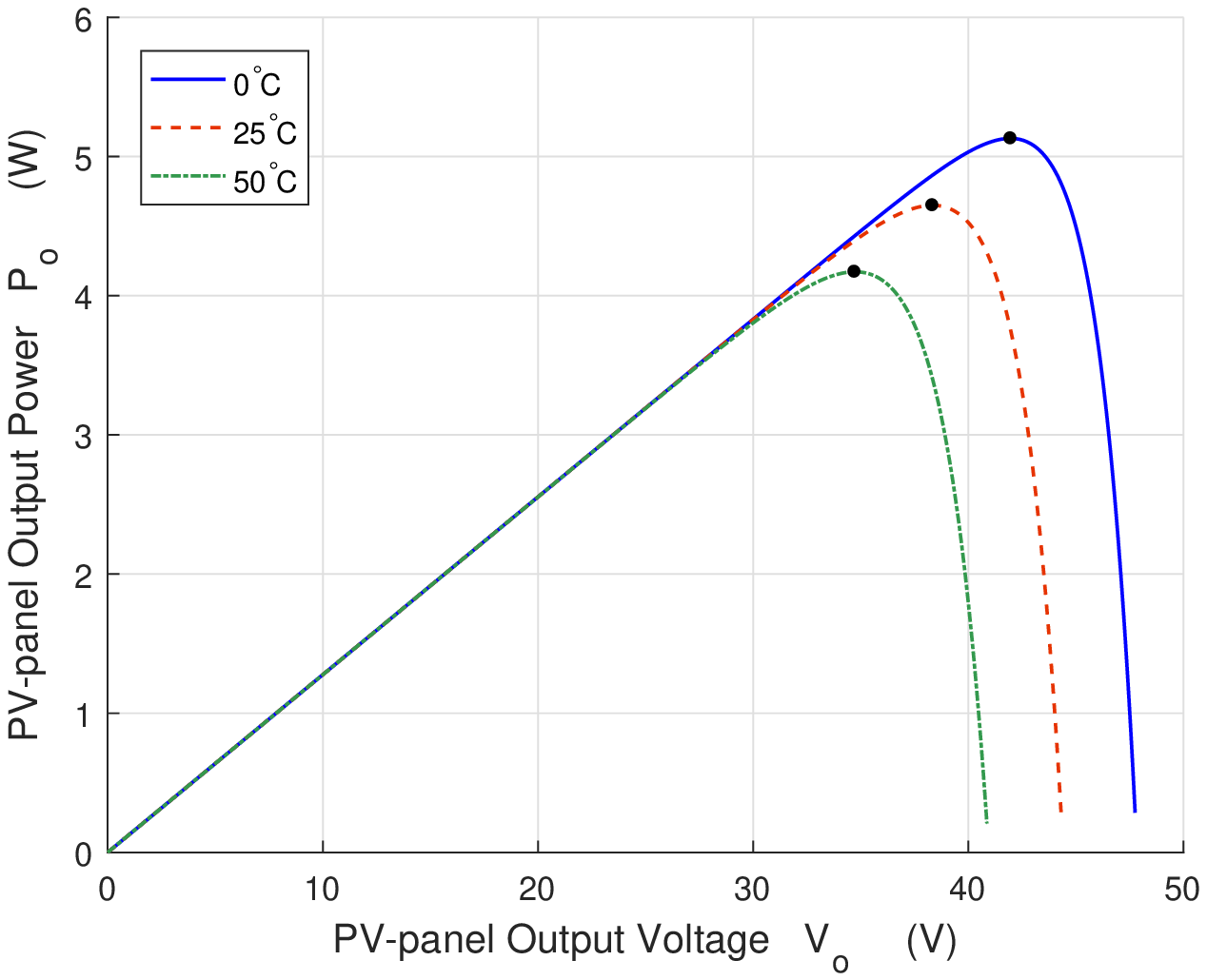}
	\caption{PV-panel Output Power vs. Voltage ($\lambda$ = 1550nm)}
    \label{1550Ptemperature}
%
	\centering
    \includegraphics[scale=0.6]{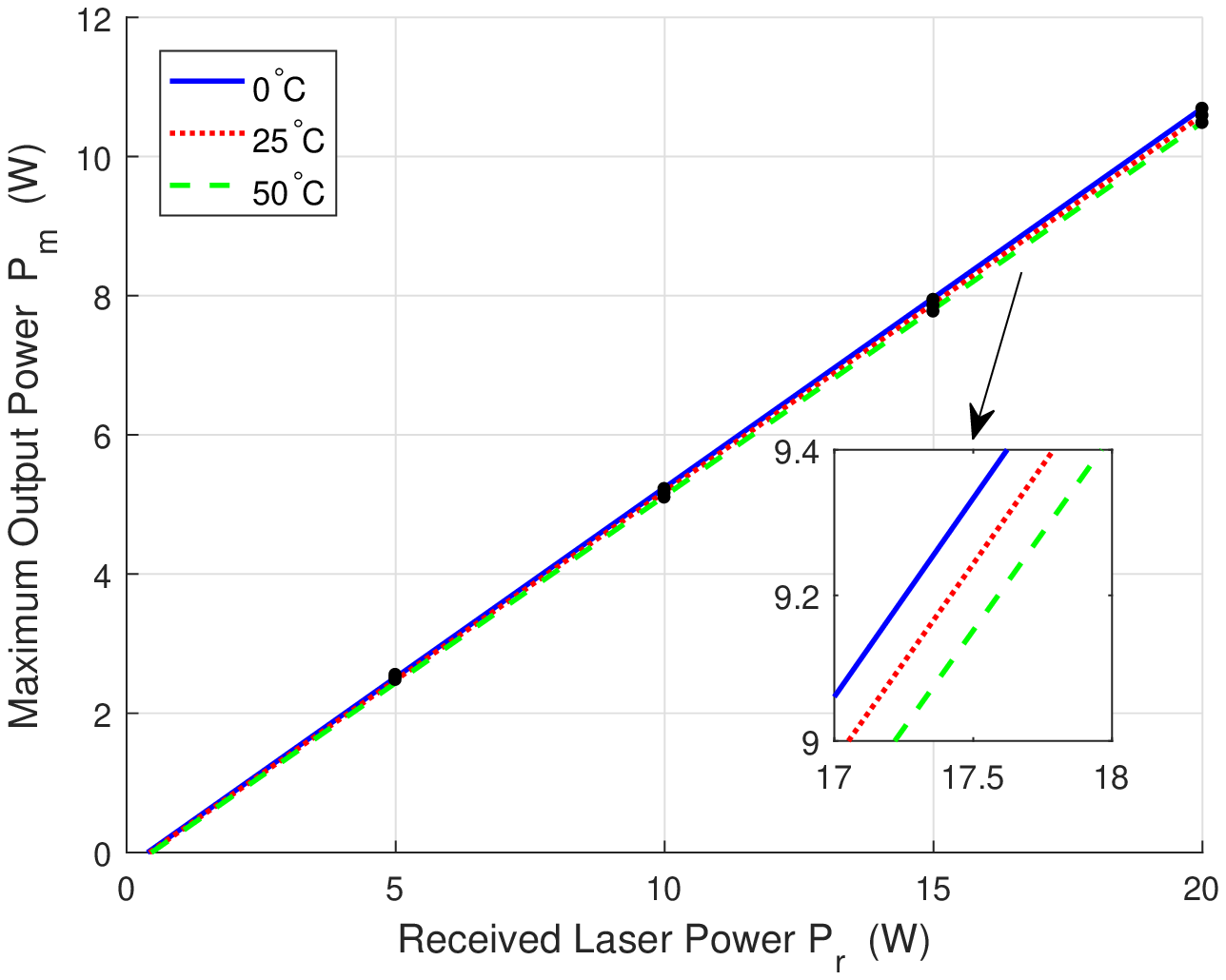}
	\caption{Maximum Output Power vs. Received Laser Power ($\lambda$ = 810nm)}
    \label{810prpm}
\end{figure}

At the DLC receiver, PV-panel takes the role of converting laser power to electrical power. PV-panel conversion efficiency relies on laser power, wavelength, and cell temperature. With reference to \eqref{Io}-\eqref{Vm}, we can obtain the PV-panel output current, voltage, and thus power, given the parameters listed in Table \ref{pvparamaters}. Fig.~\ref{810Iirradiance}-\ref{1550Ptemperature} demonstrate their relationships for different laser wavelength using the standard \emph{solar cell} Simulink model \cite{solarcell}.

Fig.~\ref{810Iirradiance} shows the relationship between PV-panel output current $I_{o}$ and voltage $V_{o}$ with different input laser power, i.e., the external-cavity laser power $P_{r}$ at the receiver, for the GaAs-based PV-panel with 810nm laser at 25$^{\circ}$C \cite{810nmpv}. Similarly, Fig.~\ref{1550Iirradiance} is for the GaSb-based PV-panel with 1550nm laser at 25$^{\circ}$C \cite{1550nmpv}. The PV-panel output power $P_{o}$ can be derived from the corresponding $I_{o}$ and $V_{o}$ based on Fig.~\ref{810Iirradiance} and Fig.~\ref{1550Iirradiance}. Thus, Fig.~ \ref{810Pirradiance} and Fig.~\ref{1550Pirradiance} depict the relationship between $P_{o}$ and $V_{o}$ for 810nm and 1550nm, respectively.

From Fig.~\ref{810Pirradiance} and Fig.~\ref{1550Pirradiance}, given $P_{r}$, we can figure out the maximum output power, which is defined as the maximum power point (MPP) and marked by the dots on the corresponding output power curves. We denote $P_m$ as the MPP of $P_o$. From \cite{onlyMPP2}, $P_m$ is proved as the unique output power, i.e., the corresponding current and voltage are unique, given the received laser power $P_r$. For example, given $P_{r} = 10$W, the MPP is unique as 4.64W for 1550nm, which is depicted by the dots in Fig.~\ref{1550Iirradiance} and \ref{1550Pirradiance}. The corresponding unique $I_{o}$ and $V_{o}$ are 121.3mA and 38.3V, respectively.

In Fig.~\ref{810Iirradiance} and Fig.~\ref{1550Iirradiance}, given $P_{r}$, $I_{o}$ keeps almost a constant when $V_{o}$ is below the MPP. However, $I_{o}$ drops rapidly when $V_{o}$ is over the MPP. For the same $V_{o}$, $I_{o}$ increases when $P_{r}$ increases. When $I_{o}$ is close to zero, $V_{o}$ is the open-circuit voltage, which increases when $P_{r}$ increases. From Fig.~\ref{810Pirradiance} and Fig.~\ref{1550Pirradiance}, given $P_{r}$, $P_{o}$ increases when $V_{o}$ increases until it reaches the MPP. However, $P_{o}$ drops dramatically when $V_{o}$ is above the corresponding voltage for MPP. For a given voltage $V_{o}$, the output power $P_{o}$ increases when the input laser power $P_{r}$ increases.

Besides input laser power, PV-cell temperature also impacts the PV-panel output current, voltage, and power. Given the three cell temperatures (0$^{\circ}$C, 25$^{\circ}$C, 50$^{\circ}$C), for $\lambda$ = 810nm and $P_r$ = 10W power, Fig.~\ref{810Itemperature} and Fig.~\ref{810Ptemperature} depict the variation of $I_{o}$ and $P_{o}$ on different $V_{o}$, respectively. Similarly, for $\lambda$ = 1550nm and $P_r$ = 10W power, Fig.~\ref{1550Itemperature} and Fig.~\ref{1550Ptemperature} show the PV-panel output $I_{o}$, $V_{o}$ and $P_{o}$ for these cell temperatures.

\begin{figure}
	\centering
    \includegraphics[scale=0.6]{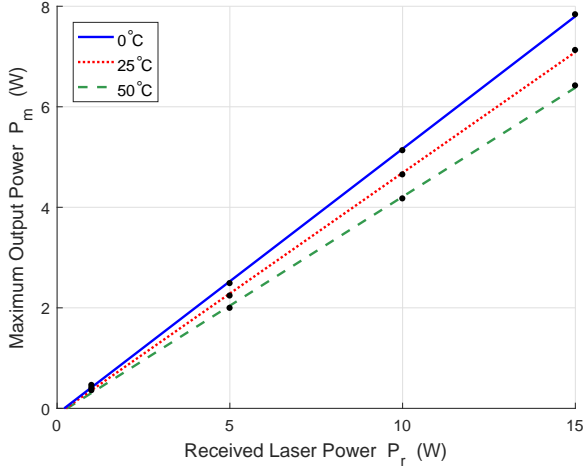}
	\caption{Maximum Output Power vs. Received Laser Power ($\lambda$ = 1550nm)}
    \label{1550prpm}
\end{figure}

From Fig.~\ref{810Itemperature} and Fig.~\ref{1550Itemperature}, $I_{o}$ keeps almost as a constant when $V_{o}$ is below a certain value. Given different cell temperatures, $I_{o}$ curves start dropping at different $V_{o}$. The turning voltage is low when the temperature is high. From Fig.~\ref{810Ptemperature} and Fig.~\ref{1550Ptemperature}, $P_{o}$ is low when the temperature is high. Additionally, the MPP increases as the cell temperature declines.

\begin{figure}
	\centering
    \includegraphics[scale=0.6]{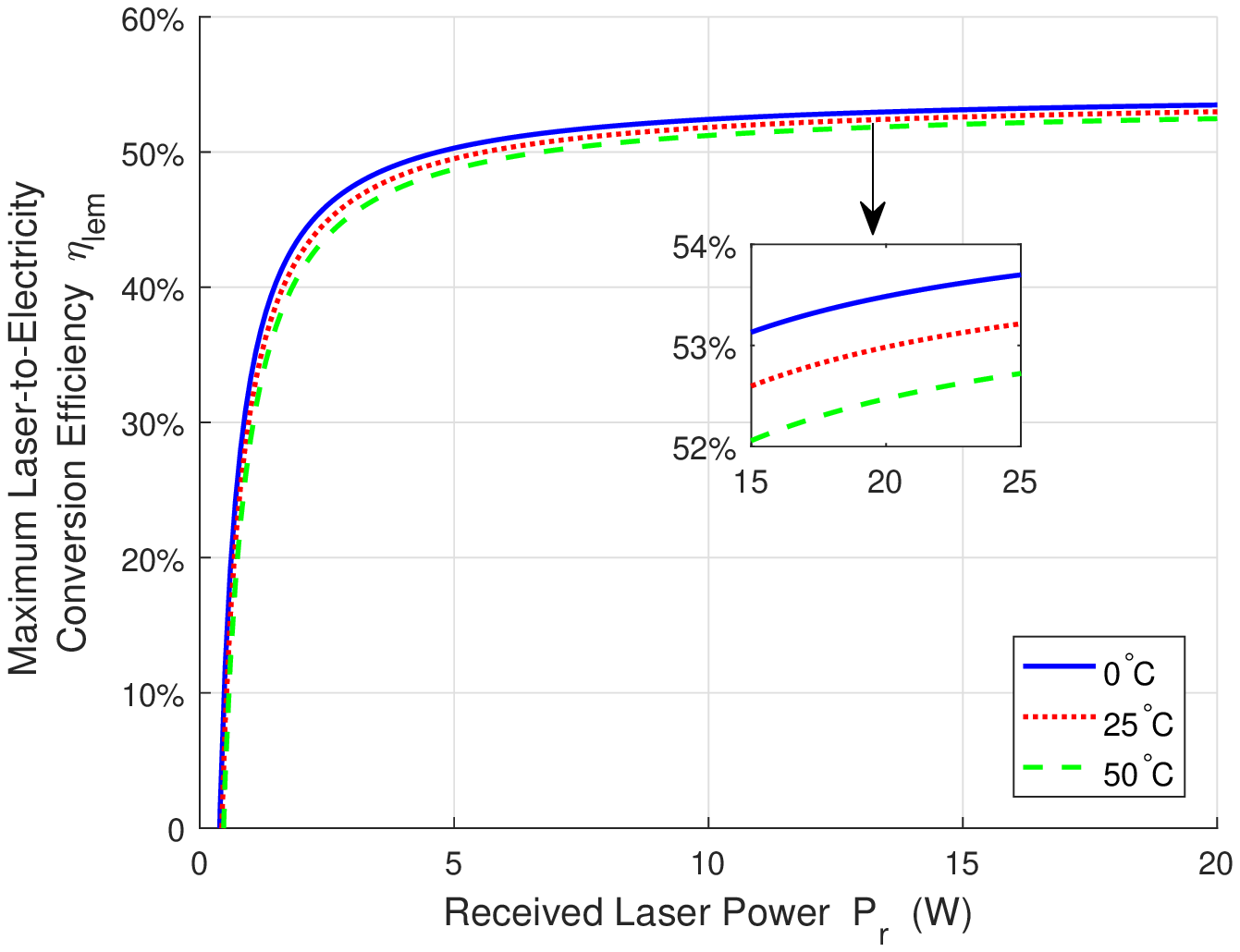}
	\caption{Maximum Laser-to-Electricity Conversion Efficiency vs. Received Laser Power ($\lambda$ = 810nm)}
    \label{810etale}
\end{figure}

\begin{figure}
	\centering
    \includegraphics[scale=0.6]{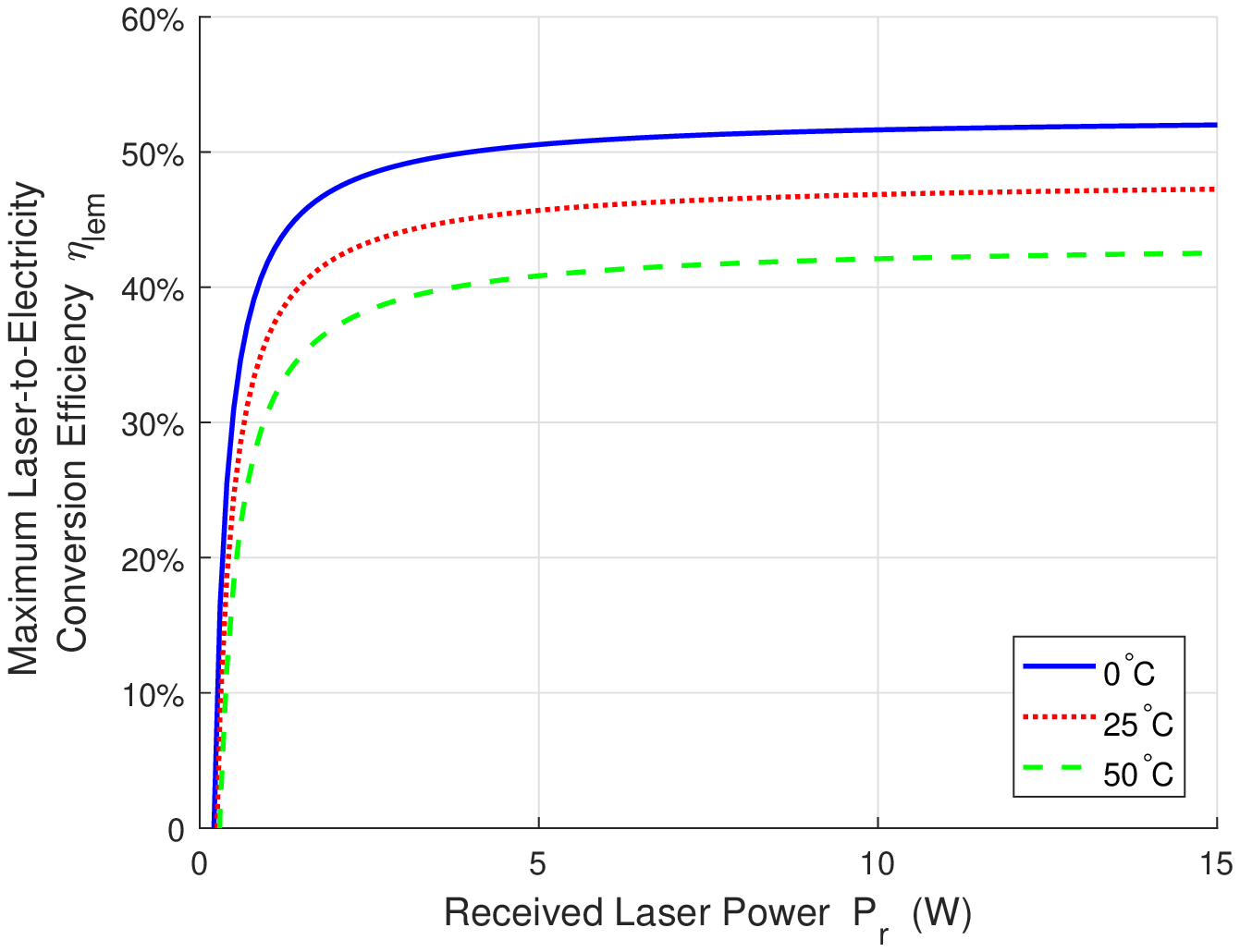}
	\caption{Maximum Laser-to-Electricity Conversion Efficiency vs. Received Laser Power ($\lambda$ = 1550nm)}
    \label{1550etale}
\end{figure}

Based on the MPP dots in Fig.~\ref{810Iirradiance} and Fig.~\ref{810Pirradiance} for different $P_r$ and Fig.~\ref{810Itemperature} and Fig.~\ref{810Ptemperature} for different cell temperatures, we can obtain the MPP dots in Fig.~\ref{810prpm}, which illustrates $P_m$ versus $P_r$ for 810nm. Similarly, Fig.~\ref{1550prpm} demonstrates $P_m$ versus $P_r$ for 1550nm. In order to evaluate the relationship between $P_m$ and $P_r$, we adopt the approximation formula by using the curve fitting method as:
\begin{equation}\label{prpmf}
  P_{m} \approx a_2 P_{r} + b_2,
\end{equation}
where $a_2$ and $b_2$ are the linear curve fitting coefficients for different wavelengths and cell temperatures, which are listed in Table \ref{pvparamaters}. From Fig.~\ref{810prpm} and Fig.~\ref{1550prpm}, we can find that the approximate lines based on \eqref{prpmf} matches the MPP dots very well.

We denote $\eta_{lem}$ as the maximum PV-panel conversion efficiency when $P_o$ is $P_m$. Based on \eqref{etapv} and \eqref{prpmf}, $\eta_{lem}$ can be depicted as:
\begin{equation}\label{etalem}
  \eta_{lem} = \frac{P_{m}}{P_{r}}= a_2+\frac{b_2}{P_{r}}.
\end{equation}

\begin{figure}
	\centering
    \includegraphics[scale=0.6]{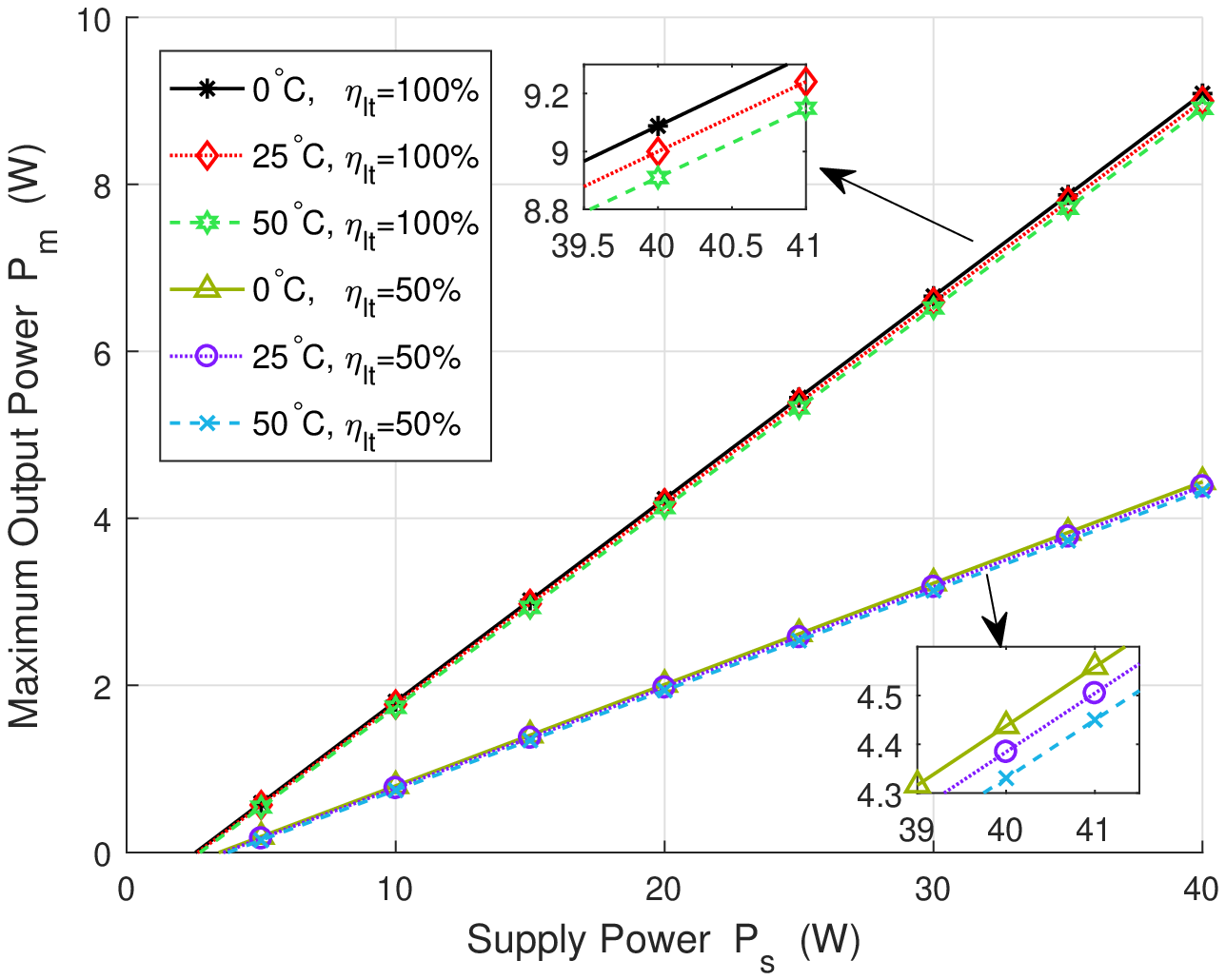}
	\caption{Maximum Output Power vs. Supply Power ($\lambda$ = 810nm)}
    \label{810pspm}
%
	\centering
    \includegraphics[scale=0.6]{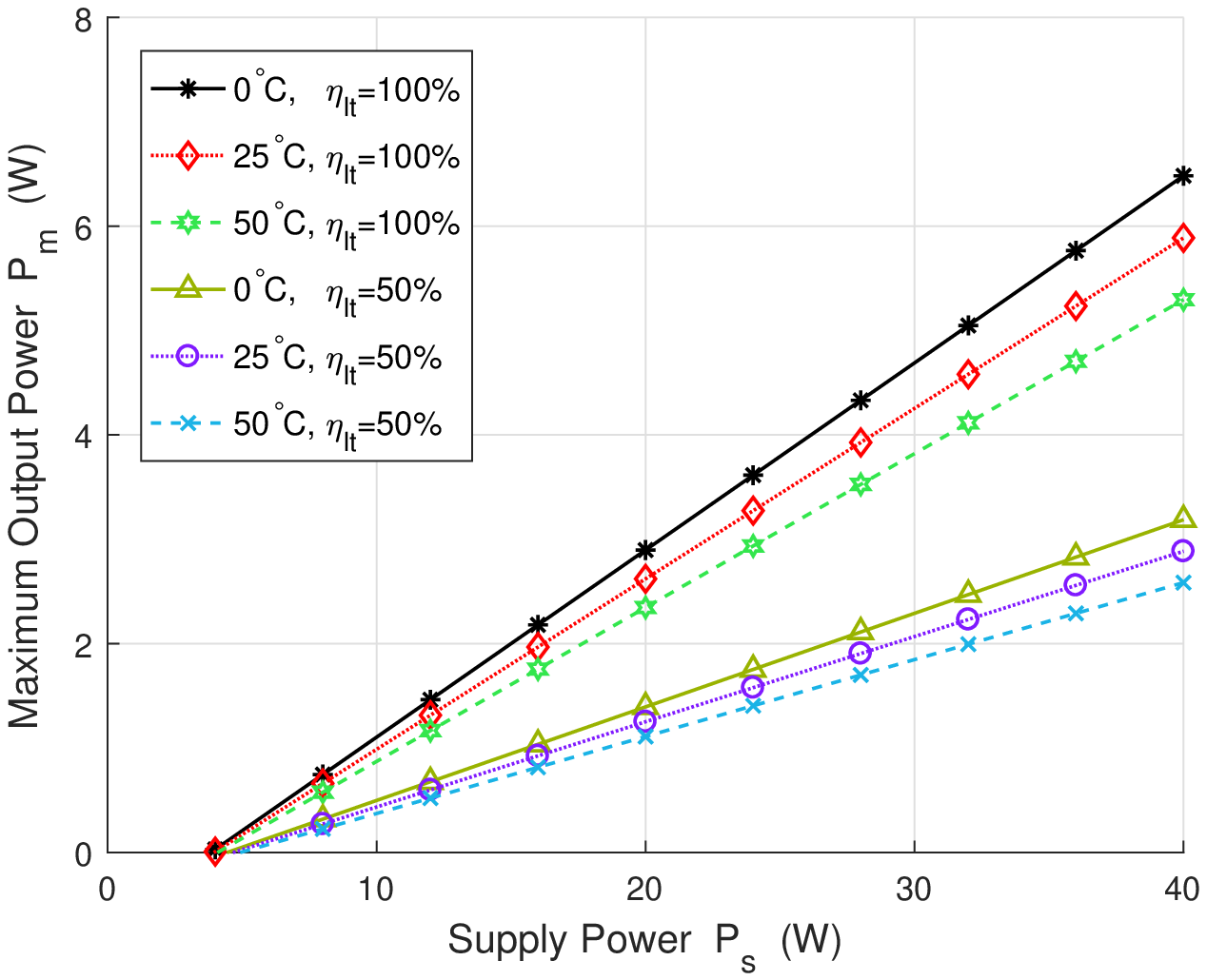}
	\caption{Maximum Output Power vs. Supply Power ($\lambda$ = 1550nm)}
    \label{1550pspm}
	\centering
    \includegraphics[scale=0.6]{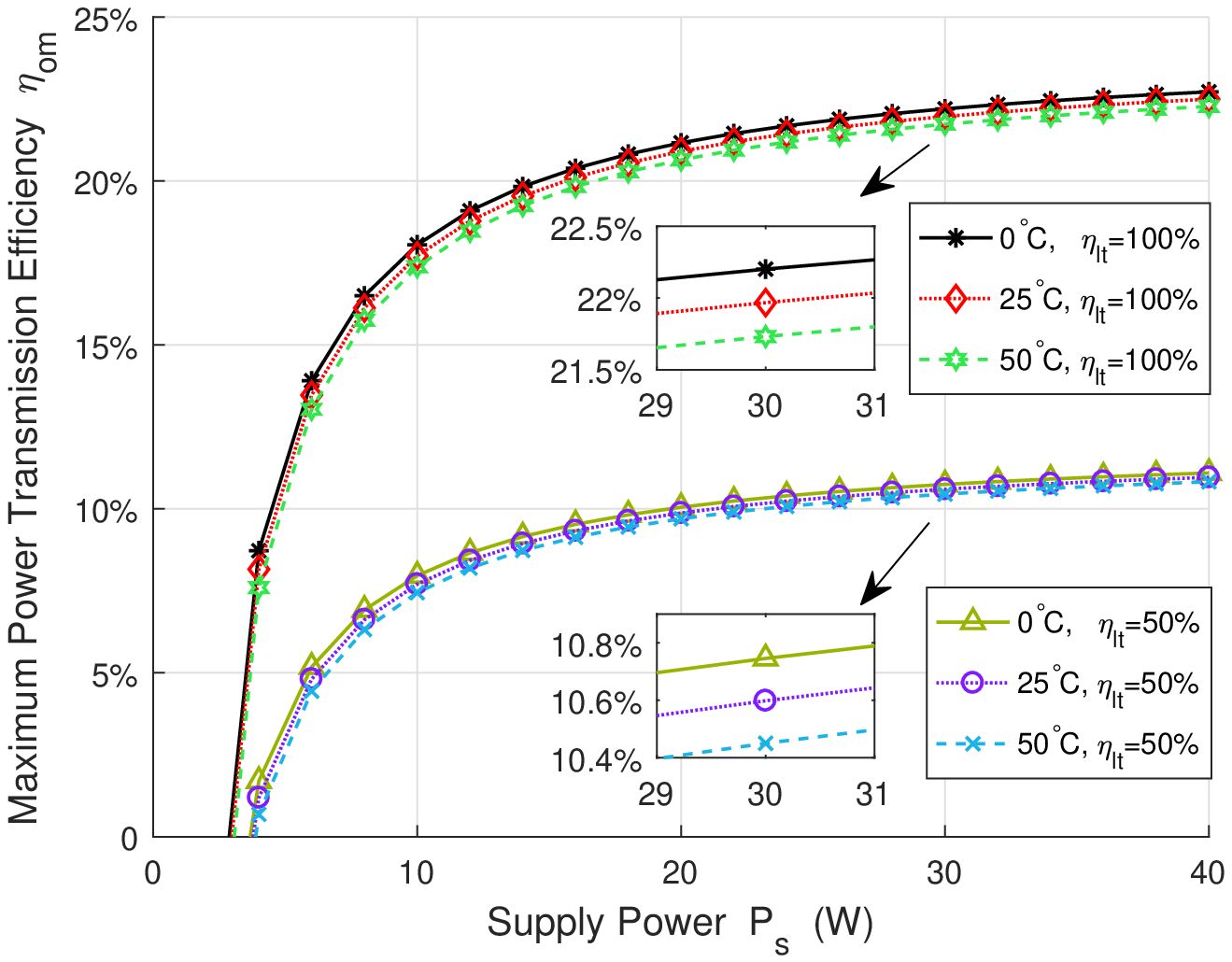}
	\caption{Maximum Power Transmission Efficiency vs. Supply Power ($\lambda$ = 810nm)}
    \label{810etao}
\end{figure}

\begin{figure}
	\centering
    \includegraphics[scale=0.6]{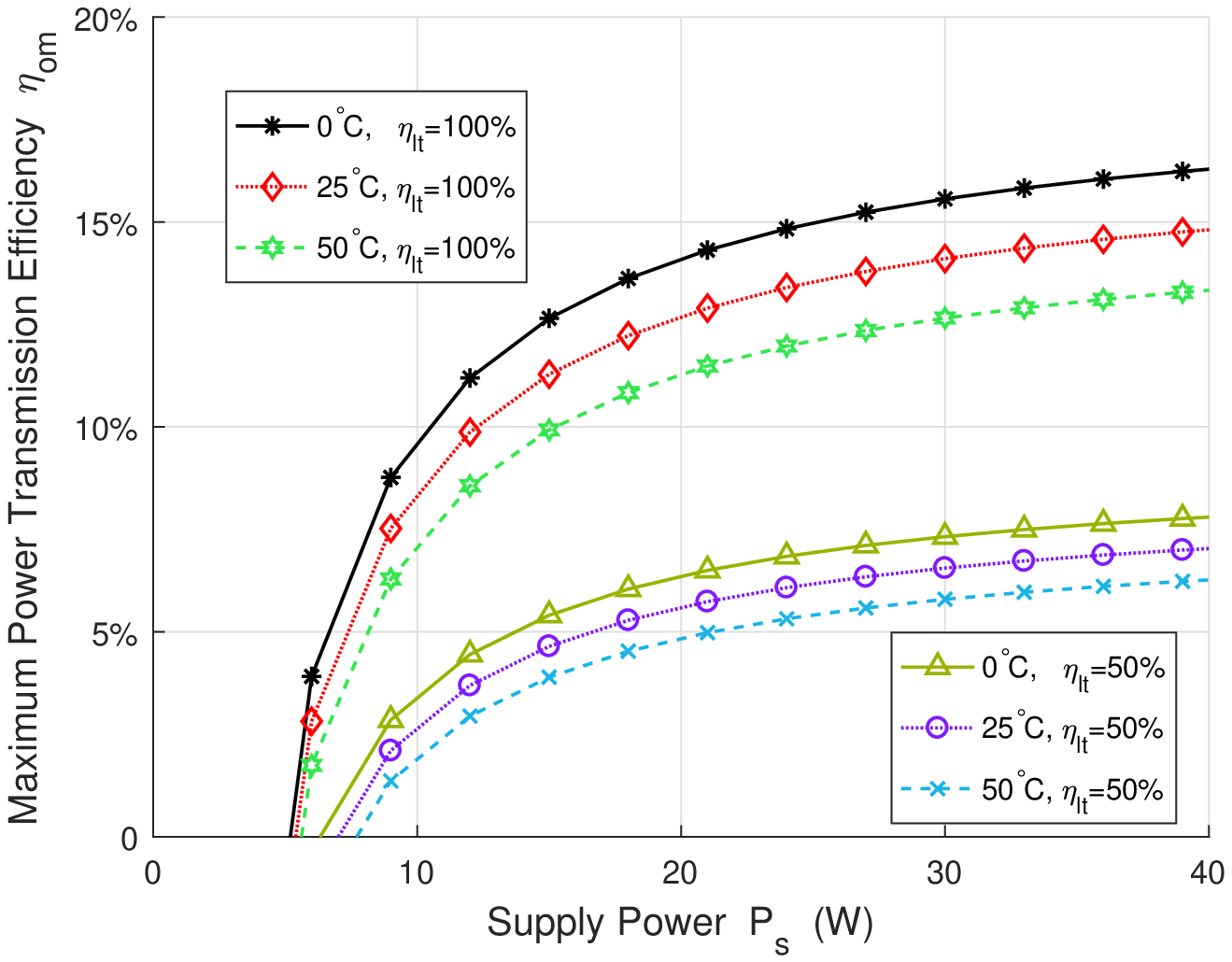}
	\caption{Maximum Power Transmission Efficiency vs. Supply Power ($\lambda$ = 1550nm)}
    \label{1550etao}
	\centering
    \includegraphics[scale=0.6]{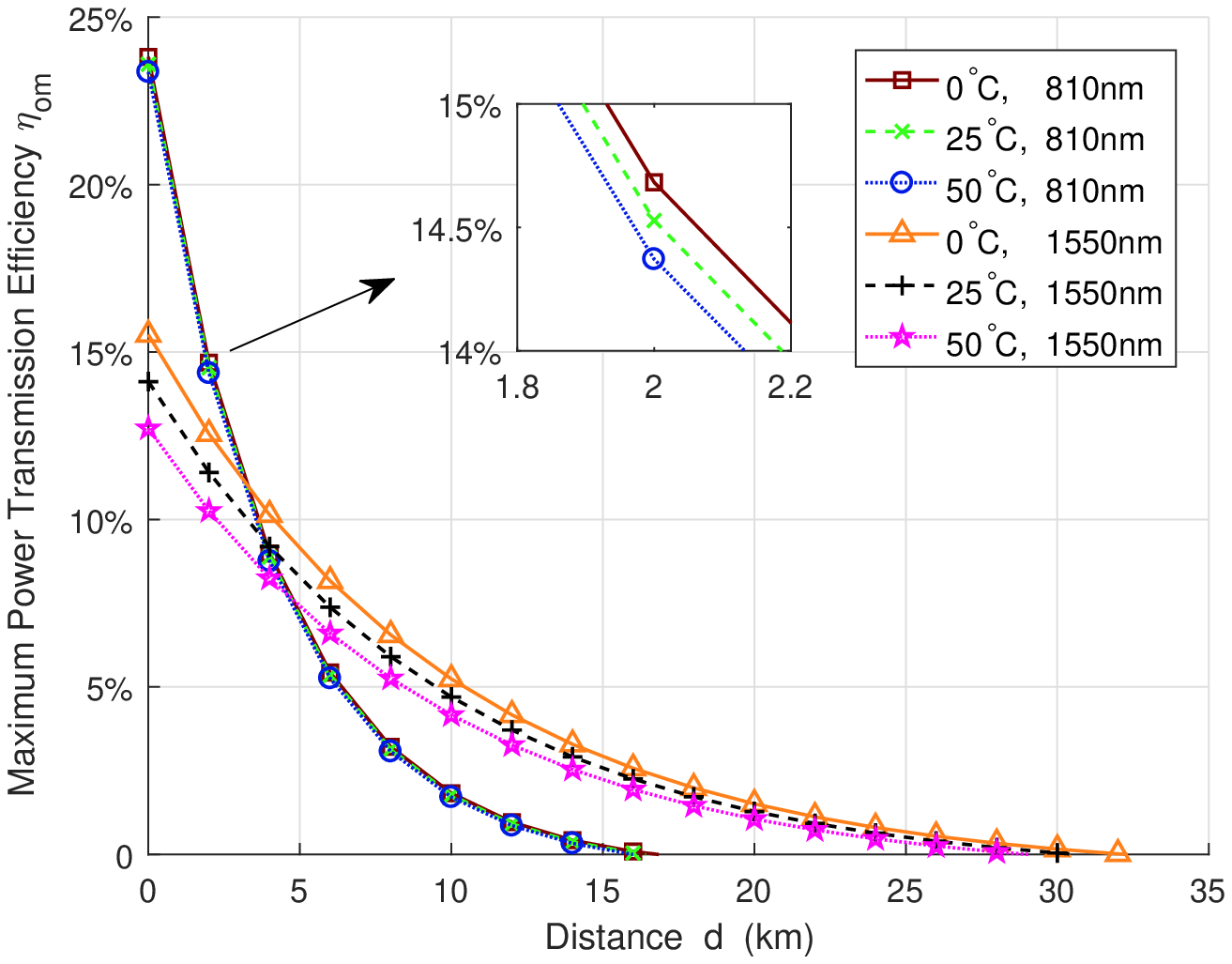}
	\caption{Maximum Power Transmission Efficiency vs. Distance (Clear Air)}
    \label{etaodc}
	\centering
    \includegraphics[scale=0.6]{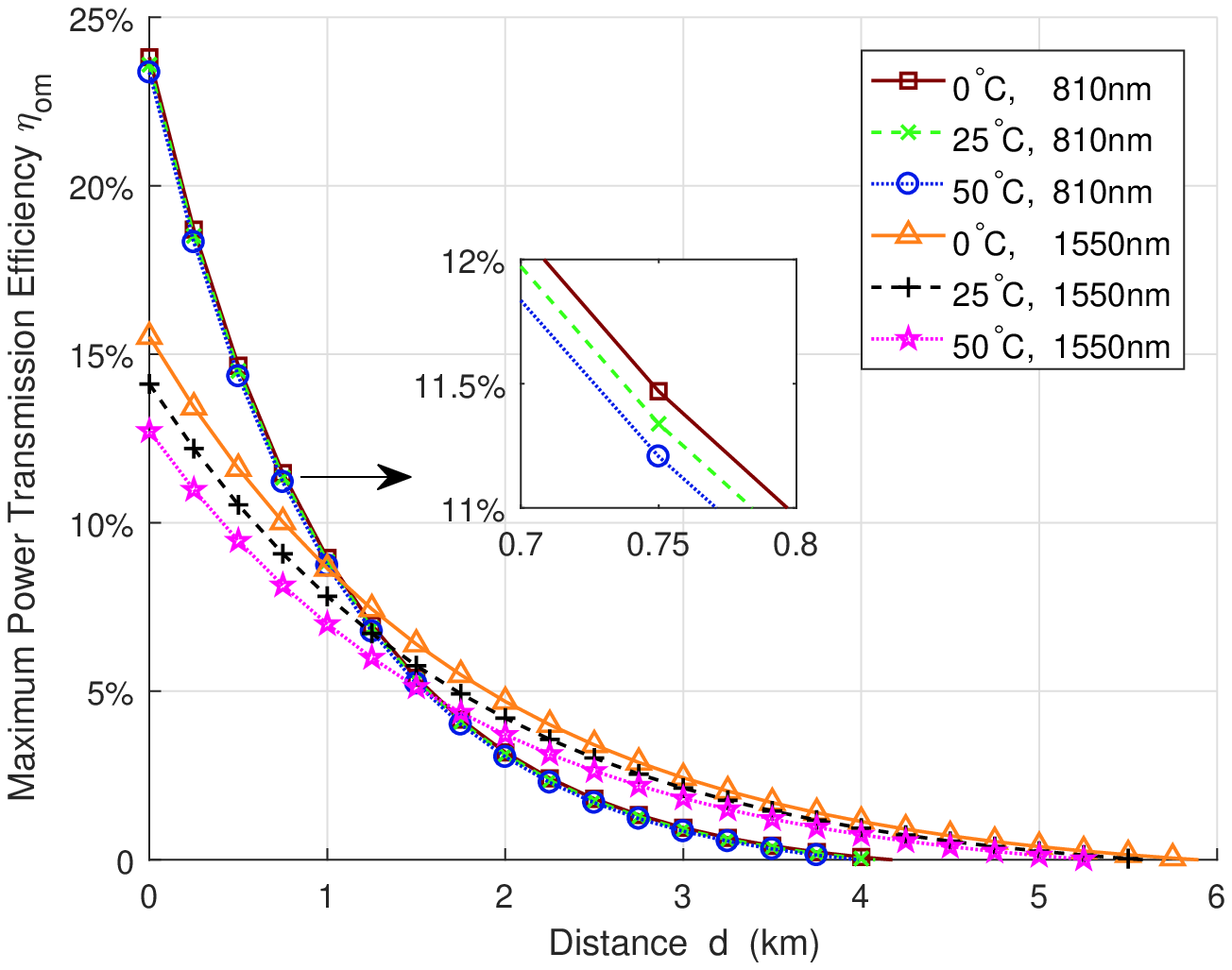}
	\caption{Maximum Power Transmission Efficiency vs. Distance (Haze)}
    \label{etaodh}
\end{figure}

Fig.~\ref{810etale} and Fig.~\ref{1550etale} show how $\eta_{lem}$ varies with the received laser power $P_{r}$ for 810nm and 1550nm, respectively. From Fig.~\ref{810etale} and Fig.~\ref{1550etale}, the changing trend of $\eta_{lem}$ is similar with that of $\eta_{el}$ in Fig.~\ref{8101550etael}. 
$\eta_{lem}$ is low when cell temperature is high. The impact of cell temperature on $\eta_{lem}$ is bigger for 1550nm than that of 810nm, comparing Fig.~\ref{810etale} and Fig.~\ref{1550etale}.

\subsection{DLC Power Transmission Efficiency}\label{}

The relationship between the laser power $P_l$ and the supply power $P_s$ is demonstrated by \eqref{plps}, when the transmission distance $d$ is close to zero in the electricity-to-laser conversion. The relationship between $P_l$ and the received laser power $P_r$ due to laser transmission is illustrated in \eqref{etalt}. The relationship between $P_r$ and the maximum PV-panel output power $P_m$ in the laser-to-electricity conversion is shown in \eqref{prpmf}. Thus, from \eqref{plps}, \eqref{etalt} and \eqref{prpmf}, we can obtain the relationship between $P_s$ at the transmitter and $P_m$ at the receiver as:

\begin{equation}\label{gspspm}
\begin{aligned}
  P_{m} = a_2 \eta_{lt} P_l+ b_2 \qquad\qquad\qquad\  \\
  = a_1 a_2 \eta_{lt} P_{s} + (a_2 b_{1} \eta_{lt} + b_2).
\end{aligned}
\end{equation}

Fig.~\ref{810pspm} depicts the linear relationship between $P_m$ and $P_s$ for $\eta_{lt}$ = 100\% and $\eta_{lt}$ = 50\%, respectively, when PV-cell temperature is 0$^{\circ}$C, 25$^{\circ}$C, 50$^{\circ}$C, and $\lambda$ = 810nm. Meanwhile, Fig.~\ref{1550pspm} illustrates the similar circumstances for $\lambda$ = 1550nm.

From \eqref{etao}, we denote $\eta_{om}$ as the maximum power transmission efficiency, when $P_{o}$ is $P_{m}$, i.e. $\eta_{le}$ approaches $\eta_{lem}$. From \eqref{etael}, \eqref{etalt}, \eqref{etapv}, \eqref{etao}, \eqref{etaelall} and \eqref{etalem}, the maximum power transmission efficiency $\eta_{lem}$ can be obtained as:
\begin{equation}\label{etaoall}
\begin{aligned}
  \eta_{om} 
  &=\eta_{el}\eta_{lt}\eta_{lem} \qquad\qquad\quad   \\
  &=\eta_{el}\eta_{lt}(a_2+\frac{b_2}{\eta_{el}\eta_{lt}P_{s}}) \\
  &=a_1 a_2 \eta_{lt} + \frac{a_2 b_{1} \eta_{lt} + b_2}{P_{s}} \\
  &=a_1 a_2 e^{-\alpha d} + \frac{a_2 b_{1} e^{-\alpha d} + b_2}{P_{s}}.
\end{aligned}
\end{equation}

\begin{figure}
	\centering
    \includegraphics[scale=0.6]{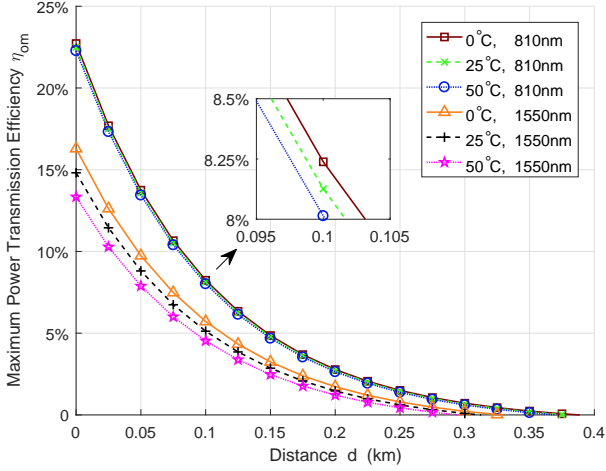}
	\caption{Maximum Power Transmission Efficiency vs. Distance (Fog)}
    \label{etaodf}
\end{figure}

Fig.~\ref{810etao} shows the relationship between $\eta_{om}$ and $P_{s}$ when $\eta_{lt}$ are 100\% and 50\% and cell temperatures are 0$^{\circ}$C, 25$^{\circ}$C, and 50$^{\circ}$C for 810nm. Fig.~\ref{1550etao} shows the same circumstances for 1550nm. $\eta_{om}$ raises up with $P_{s}$ increasing at first, then it reaches the plateau. The growth pattern of $\eta_{om}$ in Fig.~\ref{810etao} and Fig.~\ref{1550etao} is similar as $\eta_{el}$ in Fig.~\ref{8101550etael} and $\eta_{lem}$ in Fig.~\ref{810etale} and Fig.~\ref{1550etale}.

\begin{figure}
	\centering
    \includegraphics[scale=0.6]{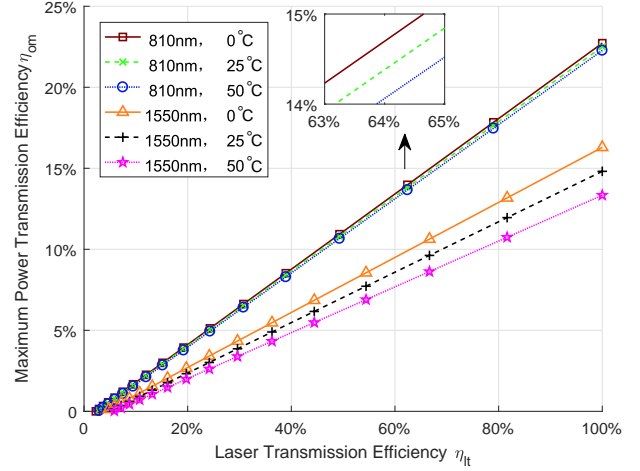}
	\caption{Maximum Power Transmission Efficiency vs. Laser Transmission Efficiency}
    \label{etaopl}
\end{figure}

$\eta_{om}$ depends not only on the supply power $P_{s}$ but also on the distance $d$. Fig.~\ref{etaodc} depicts the relationship between $\eta_{om}$ and $d$ for different laser wavelength and PV-cell temperature, when $P_{s}$ = 40W and air quality is clear. Fig.~\ref{etaodh} and Fig.~\ref{etaodf} illustrate $\eta_{om}$ for the similar situation when air condition is haze and fog, respectively. Fig.~\ref{etaopl} describes how $\eta_{om}$ changes over $\eta_{lt}$ under clear air when $P_{s}$ is 40W.

From Fig.~\ref{etaodc} and Fig.~\ref{etaodh}, $\eta_{om}$ decreases when $d$ increases. $\eta_{om}$ of 810nm laser is higher than that of 1550nm laser when $d$ is short. However, $\eta_{om}$ for 810nm is lower than that of 1550nm when $d$ is long. From Fig.~\ref{etaodf}, $\eta_{om}$ of 810nm laser always keep higher than that of 1550nm laser until $\eta_{om}$ decrease to 0. At the same time, as described above, the cell temperature has bigger impact on $\eta_{om}$ for 1550nm than that of 810nm.

From Fig.~\ref{etaopl}, $\eta_{om}$ increases linearly as $\eta_{lt}$ enhances based on \eqref{etaoall}. Fig.~\ref{etaopl} provides a guideline of designing the DLC systems. For example, if 20\% of DLC maximum transmission efficiency is expected, the 1550nm DLC system can not meet the requirement, however, the 810nm DLC system is preferred. Meanwhile, when deploying the DLC system, the transmission efficiency at a certain distance provides the theoretical reference to determine the radius, i.e., the coverage, which is similar to the base station coverage analysis in mobile communications \cite{zhou2008energy,Erchin2007novel}. Therefore, the maximum economic benefits can be obtained by minimizing the number of DLC transmitters to cover a given area \cite{gtzhou2010}. This analysis provides a guideline for the efficient deployment of the DLC systems.

In summary, the numerical evaluation in this section validates the analytical model presented in Section \ref{Section3}. At first, for the three modules: electricity-to-laser conversion, laser transmission, laser-to-electricity conversion, the conversion or transmission efficiency of each module is quantitatively analyzed. Secondly, through numerical analysis, we obtain the approximate linear relationship between the supply power $P_s$ at the transmitter and the maximum PV-panel output power $P_m$ at the receiver. Next, the maximum DLC power transmission efficiency $\eta_{om}$ in closed-form is derived. Finally, based on the maximum power transmission efficiency, DLC system design and development guidelines are provided, for example, how to select the laser wavelength and determine the coverage of the DLC systems.

\section{Conclusions}\label{Section5}

This paper presents the distributed laser charging technology for wireless power transfer. The multi-module analytical modeling of distributed laser charging provides the in-depth view of its physical mechanism and mathematical formulation. The numerical evaluation illustrates the power conversion or transmission in each module under the impacts of laser wavelength, transmission attenuation, and PV-cell temperature. The linear approximation is adopted and validated by measurement and simulation for electricity-to-laser and laser-to-electricity power conversion. Thus the maximum power transmission efficiency in closed-form is derived and its performance depending on the supply power, laser wavelength, transmission distance, and PV-cell temperature is illustrated by figures. Therefore, this paper not only provides the theoretical insight, but also offers the practical guideline in system design and deployment of distributed laser charging.

Due to the space limitation, there are serval important issues unaddressed in this paper and left for our future work, some of which are briefly discussed here:

\begin{itemize}
  \item The PV-panel efficiency used in the DLC system is about 50\%, which is not much efficient. More studies on the PV-panel types, the efficiency analyzation, and the total efficiency of the DLC system could be improved in the future.
  \item Only 810nm and 1550nm laser wavelengths are considered in this paper. Wider range of wavelengths can be studied to make the DLC system more universal in the future work.
  \item To convert the PV-panel output current and voltage to different preferred charging current and voltage for different applications, the circuit or device that can convert a source of direct current from one voltage level to another is worth to be discussed in the future.
  \item The point-to-point charging procedure is well illustrated in this paper. On this basis, the accessing protocols, the scheduling algorithms, the influencing factors of power conversion and transmission, and the system optimization for charging batteries adaptively in the the point-to-multiple-point wireless charging scenario should be another interesting topic to discuss.
  \item Since point-to-multiple-point wireless power transfer is naturally supported by distributed laser charging systems, the network architecture of WPT becomes an interesting research topic worthy of further investigation. The related protocols and algorithms to effectively operate WPT networks could be developed, e.g., WPT network access protocols, WPT scheduling algorithms and so on \cite{wirelesstechniques}.
  \item It is interesting to investigate the potential simultaneous wireless information and power transfer (SWIPT) in distributed laser charing systems. Due to the huge available bandwidth conveyed by laser, such high-power and high-rate SWIPT systems have the potential to support the demanding applications, e.g., IoT, mobile virtual reality/augmented reality, ultra-high-definition video streaming and so on \cite{zhangruimimo}.

\end{itemize}

\bibliographystyle{IEEEtran}
\bibliographystyle{unsrt}
\bibliography{references}

\begin{thebibliography}{10}
\providecommand{\url}[1]{#1}
\csname url@samestyle\endcsname
\providecommand{\newblock}{\relax}
\providecommand{\bibinfo}[2]{#2}
\providecommand{\BIBentrySTDinterwordspacing}{\spaceskip=0pt\relax}
\providecommand{\BIBentryALTinterwordstretchfactor}{4}
\providecommand{\BIBentryALTinterwordspacing}{\spaceskip=\fontdimen2\font plus
\BIBentryALTinterwordstretchfactor\fontdimen3\font minus
  \fontdimen4\font\relax}
\providecommand{\BIBforeignlanguage}[2]{{%
\expandafter\ifx\csname l@#1\endcsname\relax
\typeout{** WARNING: IEEEtran.bst: No hyphenation pattern has been}%
\typeout{** loaded for the language `#1'. Using the pattern for}%
\typeout{** the default language instead.}%
\else
\language=\csname l@#1\endcsname
\fi
#2}}
\providecommand{\BIBdecl}{\relax}
\BIBdecl

\bibitem{wirelesstechniques}
X.~Lu, D.~Niyato, P.~Wang, D.~I. Kim, and Z.~Han, ``Wireless charger networking
  for mobile devices: Fundamentals, standards, and applications,'' \emph{IEEE
  Wireless Communications}, vol.~22, no.~2, pp. 126--135, 2015.

\bibitem{electromagnetic}
A.~Costanzo, M.~Dionigi, D.~Masotti, M.~Mongiardo, G.~Monti, L.~Tarricone, and
  R.~Sorrentino, ``Electromagnetic energy harvesting and wireless power
  transmission: A unified approach,'' \emph{Proceedings of the IEEE}, vol. 102,
  no.~11, pp. 1692--1711, 2014.

\bibitem{liu2016dlc}
Q.~Liu, J.~Wu, P.~Xia, S.~Zhao, W.~Chen, Y.~Yang, and L.~Hanzo, ``Charging
  unplugged: Will distributed laser charging for mobile wireless power transfer
  work?'' \emph{IEEE Vehcular Technology Magzine}, vol.~11, no.~4, pp. 36--45,
  Dec. 2016.

\bibitem{wi-charge}
W.-C.~L. Technology, ``To power with light,'' [Onilne]. Available:
  \url{http://www.wi-charge.com/}.

\bibitem{niu2013optimal}
J.~Gong, S.~Zhou, and Z.~Niu, ``Optimal power allocation for energy harvesting
  and power grid coexisting wireless communication systems,'' \emph{IEEE
  Transactions on Communications}, vol.~61, no.~7, pp. 3040--3049, 2013.

\bibitem{anna2003opportunistic}
A.~Scaglione and Y.-W. Hong, ``Opportunistic large arrays: Cooperative
  transmission in wireless multihop ad hoc networks to reach far distances,''
  \emph{IEEE Tansactions on Signal Processing}, vol.~51, no.~8, pp. 2082--2092,
  2003.

\bibitem{zhou2004performance}
S.~Zhou, Z.~Wang, and G.~B. Giannakis, ``Performance analysis for
  transmit-beamforming with finite-rate feedback,'' in \emph{Proc. of 38th
  Conf. on Info. Sciences and Systems}, 2004, pp. 17--19.

\bibitem{zhangruimimo}
R.~Zhang and C.~K. Ho, ``{MIMO} broadcasting for simultaneous wireless
  information and power transfer,'' \emph{IEEE Transactions on Wireless
  Communications}, vol.~12, no.~5, pp. 1989--2001, 2013.

\bibitem{810nmtransmitter}
L.~D. S. C. .~M. Technologies, ``Laser diode source - 808nm,'' [Onilne].
  Available:
  \url{https://www.laserdiodesource.com/shop/808nm-25Watt-Laser-Diode-Module-BWT-Beijing}.

\bibitem{1550nmtransmitter}
L.~D.~S. Seminex, ``Laser diode source - 1550nm,'' [Onilne]. Available:
  \url{https://www.laserdiodesource.com/laser-diode-product-page/1470nm-1532nm-1550nm-50W-multi-chip-fiber-coupled-module-Seminex}.

\bibitem{laserenergy2009}
L.~Summerer and O.~Purcell, ``Concepts for wireless energy transmission via
  laser,'' \emph{Europeans Space Agency Advanced Concepts Team}, 2009.

\bibitem{green2015solar}
A.~G. Martin, E.~Keith, H.~Yoshihiro, W.~Wilhelm, and D.~D. Ewan, ``Solar cell
  efficiency tables (version 45),'' \emph{Progress in Photovoltaics: Research
  and Applications}, vol.~23, no.~1, pp. 1--9, 2015.

\bibitem{laserdiodes}
B.~V.~Zeghbroeck, \emph{Principles of semiconductor devices}, 1st~ed.\hskip 1em
  plus 0.5em minus 0.4em\relax University of Colorado, 2004.

\bibitem{attenuation}
S.~A. Salman, J.~M. Khalel, and W.~H. Abas, ``Attenuation of infrared laser
  beam propagation in the atmosphere,'' \emph{Diala Jour}, vol.~36, pp. 2--9,
  2009.

\bibitem{JMLiuphotonic}
J.~M. Liu, \emph{Semiconductor lasers and light emitting diodes}, 1st~ed.\hskip
  1em plus 0.5em minus 0.4em\relax Institute of Optics, College of Engineering
  and Applied Science, University of Rochester, 2005.

\bibitem{foghaze}
I.~I. Kim, B.~McArthur, and E.~J. Korevaar, ``Comparison of laser beam
  propagation at 785 nm and 1550 nm in fog and haze for optical wireless
  communications,'' in \emph{proc. SPIE}, vol. 4214, 2001, pp. 26--37.

\bibitem{secondauthorPV}
M.~S. Aziz, S.~Ahmad, H.~Ijaz, H.~Akaash, and S.~Umair, ``Simulation and
  experimental investigation of the characteristics of a {PV}-harvester under
  different conditions,'' in \emph{2014 IEEE International Conference on Energy
  Systems and Policies (ICESP)}, Nov. 2014, pp. 1--8.

\bibitem{solarcell}
T.~Salmi, M.~Bouzguenda, A.~Gastli, and A.~Masmoudi, ``{MATLAB}/{S}imulink
  based modeling of photovoltaic cell,'' \emph{International Journal of
  Renewable Energy Research (IJRER)}, vol.~2, no.~2, pp. 213--218, 2012.

\bibitem{810nmpv}
E.~Oliv, F.~Dimroth, and A.~W. Bett, ``{GaAs} converters for high power
  densities of laser illumination,'' \emph{Progress in Photovoltaics: Research
  and Applications}, vol.~16, no.~4, pp. 289--295, 2008.

\bibitem{1550nmpv}
V.~P. Khvostikov, S.~V. Sorokina, F.~Y. Soldatenkov, and N.~K. Timoshina,
  ``Gasb-based photovoltaic laser-power converter for the wavelength $\lambda$
  $\approx$ 1550 nm,'' \emph{Semiconductors}, vol.~49, no.~8, pp. 1079--1082,
  Aug. 2015.

\bibitem{onlyMPP2}
S.~Liu and R.~A. Dougal, ``Dynamic multiphysics model for solar array,''
  \emph{IEEE Transactions on Energy Conversion}, vol.~17, no.~2, pp. 285--294,
  2002.

\bibitem{zhou2008energy}
Z.~Zhou, S.~Zhou, J.-H. Cui, and S.~Cui, ``Energy-efficient cooperative
  communication based on power control and selective single-relay in wireless
  sensor networks,'' \emph{IEEE Tansactions on Wireless Communications},
  vol.~7, no.~8, 2008.

\bibitem{Erchin2007novel}
K.-L. Noh, Q.~M. Chaudhari, E.~Serpedin, and B.~W. Suter, ``Novel clock phase
  offset and skew estimation using two-way timing message exchanges for
  wireless sensor networks,'' \emph{IEEE Tansactions on Communications},
  vol.~55, no.~4, pp. 766--777, 2007.

\bibitem{gtzhou2010}
Q.~Liu, W.~Zhang, X.~Ma, and G.~T. Zhou, ``{A practical amplify-and-forward
  relaying strategywith an intentional peak power limit},'' in
  \emph{International Conference on Acoustics, Speech, and Signal Processing},
  2010, pp. 2518--2521.

\end{thebibliography}

\end{document}